\documentclass[aps,pre,twocolumn]{revtex4}
\usepackage{graphics}
\usepackage{graphicx}

\begin{document}


\title{Subdiffusive L\'evy flights in quantum nonlinear Schr\"odinger lattices with algebraic power nonlinearity}


\author{Alexander~V.~Milovanov${}^{1,2}$ and Alexander~Iomin${}^{3}$}

\affiliation{${}^1$ENEA National Laboratory, Centro~Ricerche~Frascati, I-00044 Frascati, Rome, Italy}
\affiliation{${}^2$Space Research Institute, Russian Academy of Sciences, 117997 Moscow, Russia}
\affiliation{${}^3$Department of Physics, Technion$-$Israel Institute of Technology, 32000 Haifa, Israel}




\begin{abstract} We report a new result concerning the dynamics of an initially localized wave packet in quantum nonlinear Schr\"odinger lattices with a disordered potential. A class of nonlinear lattices with subquadratic power nonlinearity is considered. We show that there exists a parameter range for which an initially localized wave packet can spread along the lattice to unlimited distances, but the phenomenon is purely quantum, and is hindered in the corresponding classical lattices. The mechanism for this spreading is moreover very peculiar and assumes that the components of the wave field may form coupled states by tunneling under the topological barriers caused by multiple discontinuities in the operator space. Then these coupled states thought of as quasiparticle states can propagate to long distances on L\'evy flights with a distribution of waiting times. The overall process is subdiffusive and occurs as a competition between long-distance jumps of the quasiparticle states, on the one hand, and long-time trapping phenomena mediated by clustering of unstable modes in wave number space, on the other hand. The kinetic description of the transport, discussed in this work, is based on fractional-derivative equations allowing for both (i) non-Markovianity of the spreading process as a result of attractive interaction among the unstable modes; this interaction is then described in terms of the familiar Lennard-Jones potential; and (ii) the effect of long-range correlations in wave number space tending to introduce fast channels for the transport, the so-called ``stripes." We argue that the notion of stripes is key to understand the topological constraints behind the quantum spreading, and we involve the idea of stripy ordering to obtain {\it self-consistently} the parameters of the associated waiting-time and jump-length distributions. Finally, we predict the asymptotic laws for quantum transport and show that the relevant parameter determining these laws is the exponent of the power-law defining the type of the nonlinearity. The results, presented here, shed light on the origin of L\'evy flights in quantum nonlinear lattices with disorder.   
\end{abstract}


\maketitle

\section{Introduction} 
Waves in random systems cannot readily propagate to long distances: scattered by structural inhomogeneities on many spatial scales, they tend to form multiple standing waves at high disorder, and this effectively confines the wave process within a spatially restricted domain. The phenomenon$-$predicted by Anderson in 1958 \cite{And} and extensively studied ever since$-$has come to be known as the Anderson localization and occurs for any type of wave process, classical or quantum. 

A continued interest in the phenomena of Anderson localization was due to the direct experimental observation of the Anderson localization of visible light \cite{Maret1} and the measurement of the critical exponent of scaling theory of the localization transition \cite{Maret2}. More recently, there has been a stream of literature stimulated by Pikovsky and Shepelyansky \cite{Sh93,PS} that sought to demonstrate that the Anderson localization in random systems could be destroyed by a weak nonlinearity and that the phenomenon is thresholded in that there exists a critical strength of nonlinear interaction such that above this strength the nonlinear field can propagate across the lattice to infinitely long distances, and is Anderson localized despite these nonlinearities otherwise. 

Theoretically, the destruction of Anderson localization in nonlinear lattices has been studied in the fashion of the Gross-Pitaevskii (i.e., nonlinear Schr\"odinger) equation with disordered potential \cite{Sh93,PS,Wang,Flach,Skokos,Fishman,Iomin,Erez,EPL,PRE14}. A modified perturbation theory with regard to the strength of the nonlinear term has been developed \cite{Wang,Fishman}, and extensive numerical simulations have been carried out \cite{Flach,Skokos,Fishman}. A subdiffusive scaling for the onset spreading has been introduced and numerically measured \cite{PS,Flach}. A non-perturbative approach to the nonlinear Anderson problem has been developed based on topological approximations, using random walks and the concept of critical percolation on a Cayley tree \cite{EPL,PRE14,Chapter,DNC}. The subject has attracted additional interest recently in view of its extension to quantum dynamics \cite{Iomin16,Flach_89} and the suggestion$-$motivated by Fermi's golden rule$-$that the loss of localization in the quantum domain could be {\it not} thresholded \cite{PRE17}. 

Our purpose here is to describe the delocalizing effect of subquadratic power nonlinearity on quantum dynamics of a lattice gas in nonlinear Schr\"odinger lattices with disorder. A background for this consists in the following. (i) It has been shown \cite{EPL,PRE14,Chapter,DNC} based on a classical analysis that a power nonlinearity of the Ginzburg-Landau type (i.e., {quadratic} power nonlinearity) played a very special role in classical dynamics, and that the quadratic power case was the {\it only} power case to allow for a classical transport to long distances by a stochastic process. (ii) Also it has been shown \cite{PRE14} using a mapping procedure on a Cayley tree \cite{EPL} that {\it there is no asymptotic classical transport in the parameter range of subquadratic power nonlinearity} due to a certain type of topological constraints on trees. 

In the present work, we contest these classical results and show that quantum dynamics may allow the wave field to spread even for subquadratic power nonlinearities under certain conditions. The mechanism for this spreading uses the idea that the quantum waves may tunnel under the topological barriers associated with multiple discontinuities in the operator space. Then the waves with oppositely directed momenta can couple together to form a joint state which also interacts with the other states. The result of this interaction is formation of one-dimensional ordered structures in wave number space, which we arguably call ``stripes" \cite{PRB02}. Then the nonlinear field may propagate to long distances along these stripes, and we show that the process occurs in the form of a competition between long-time trapping phenomena due to the clustering of the unstable modes, on the one hand, and instantaneous jumps of the coupled states in random direction along the stripes, on the other hand. 

Next we show based on a simple random-walk model that the clustering phenomena in wave number space do introduce a nontrivial statistics of exit times, with a diverging mean waiting time, thus leading to important non-Markovian features with a heavy-tailed auto-correlation; whereas the jumps of the coupled states along the stripes introduce competing {\it nonlocal} features \cite{PLA05} and can be understood dynamically in terms of L\'evy flights \cite{Klafter,Klafter2004,Ch2007}. In that regard, we argue that the L\'evy flights are a characteristic of quantum models with subquadratic power nonlinearity in that they do {\it not} appear in the corresponding classical descriptions \cite{PRE14,Chapter,DNC}, nor in quantum models with the quadratic power nonlinearity \cite{PRE17}. The kinetic approach to the asymptotic transport, discussed in this work, is based on fractional derivative equations \cite{Klafter,Klafter2004,Ch2007,Nature,Sokolov} accounting for both the waiting-time statistics with a distribution of waiting times and the long-distance jumps of the coupled states along the stripes. Our results shed light on the quantum routes to {\it fractional kinetics} \cite{Klafter,Nature,Sokolov} and the quantum significance of L\'evy flights as a partial case.  

The paper is organized as follows. The quantum model is described first (Sec. II; the preamble), followed by a construction of the backbone map (Sec. II\,A) and the associated backbone-reduced dynamical equations (Sec. II\,B). In Sec. III we discuss the relevant topological methods using the concepts of a Cayley tree, a Cayley forest, and stripes. In Sec. IV we obtain the laws of spreading by solving a dynamical problem for a ``particle" interacting with a potential field of the Lennard-Jones type \cite{Lennard}. Also in Sec. IV we identify a trapping mechanism for quantum subdiffusion (in terms of clustering of unstable modes). Section V is concerned with a kinetic picture of the transport based on a fractional generalization of the diffusion equation. Section VI is a discussion session and summarizes a few elements to our approach needing a broader interpretation. In Sec. VII we explain an apparent {\it paradox} demonstrating the basic physics significance of the Anderson localization. We reiterate on our results in the concluding Sec. VIII with a few remarks.             
  
\section{The Model}
We consider the problem of dynamical localization of waves in a quantum nonlinear Schr\"odinger equation (QNLSE) with random potential, i.e.,
\begin{equation}
i\hbar\frac{\partial\hat\psi_n}{\partial t} = \hat{H}_L\hat\psi_n + \beta |\hat\psi_n|^{2s}\hat\psi_n,
\label{1} 
\end{equation}
where $\hat\psi_n = \hat\psi (n, t)$ is an operator wave function and is defined on a grid with the discrete coordinate $n$; $|\hat\psi_n|^{2s}\equiv (|\hat\psi_n|^{2})^s$ is the definition of the algebraic power nonlinearity used in this work; $s$ is a power exponent and is taken from the unit interval $0 < s\leq 1$; in this regard $s=1$ represents the familiar quadratic nonlinearity of the Ginzburg-Landau type; $s\rightarrow 0$ represents the linear localization case; $s < 1$ represents a subquadratic power nonlinearity and is the main focus of this study; further $|\hat\psi_n|^{2}\equiv \hat\psi_n^{\dag}\hat\psi_n$ is the amplitude of the wave field; $\hat\psi_n^{\dag}$ with the superscript $^{\dag}$ is Hermitian conjugate wave function; 
\begin{equation}
\hat{H}_L\hat\psi_n = \varepsilon_n\hat\psi_n + V (\hat\psi_{n+1} + \hat\psi_{n-1})
\label{2} 
\end{equation}
is the Hamiltonian of a linear problem in the tight-binding approximation \cite{And}; the coefficient $\beta$ characterizes the strength of nonlinearity; on-site energies $\varepsilon_n$ are randomly distributed with zero mean across a finite energy range; $V$ is hopping matrix element; and $\hbar$ is Planck's constant and is set to 1 hereafter. The total (summed over all $n$) probability is 1 and in the operator form corresponds to $\sum_n \hat\psi_n^{\dag}\hat\psi_n = \hat 1$, where $\hat 1$ is the unity operator. For $\beta\rightarrow 0$, QNLSE with the Hamiltonian operator in Eq.~(\ref{2}) offers a quantum representation of the linear Anderson model in Ref. \cite{And}. 

Next, the eigenstates, $\phi_{n,k}$, of the linear model are defined through $\hat H_L \phi_{n,k} = \omega_k \phi_{n,k}$ and constitute a full basis of mutually orthogonal complex functions with the eigenfrequencies $\omega_k$, where $k=0 ,\pm 1,\pm 2, \dots$ is an integer number. Note that all eigenstates $\phi_{n,k}$ are exponentially localized in the linear regime, i.e., no spreading is expected to occur for $\beta = 0$. 

To obtain the laws of spreading in the nonlinear phase, it is convenient to consider the operator wave function $\hat\psi_n$ as a ``vector" in functional space, whose basis vectors are the eigenstates of the linear problem with $\beta = 0$. Using the functions $\phi_{n,k}$ as the basis functions, we write  
\begin{equation}
\hat\psi_n = \sum_m \hat a_m (t) \phi_{n,m},
\label{Expan} 
\end{equation}
and similarly for $\hat\psi_n^{\dag}$, i.e., $\hat\psi_n^{\dag} = \sum_m \hat a_m^{\dag} (t) \phi^*_{n,m}$. Without loss in generality, we consider the eigenfunctions $\phi_{n,k}$ being normalized to unity, with the natural orthonormality condition 
\begin{equation}
\sum _n \phi^*_{n,k_1}\phi_{n,k_2} = \delta_{k_1,k_2}.
\label{Orton} 
\end{equation}
Here, $\delta_{k_1,k_2}$ is Kronecker's delta and star denotes complex conjugation. $\hat a_m (t)$ and $\hat a_m^{\dag} (t)$ are, respectively, the annihilation and the creation bosonic operators obeying the commutation rule 
\begin{equation}
[\hat a_{m_1}, \hat a_{m_2}^{\dag}] = \hat a_{m_1} \hat a_{m_2}^{\dag} - \hat a_{m_2}^{\dag} \hat a_{m_1} = \delta_{m_1,m_2}.
\label{Commut} 
\end{equation}
With the aid of Eq.~(\ref{Orton}), one sees that $[\hat \psi_{i}, \hat \psi_{j}^{\dag}] = \delta_{i,j}$ for all pairs of indices $i,j$. The total probability being equal to 1 implies 
\begin{equation}
\sum_n \hat\psi_n^{\dag}\hat\psi_n = \sum_m \hat a_m^{\dag} (t) \hat a_m (t) = \hat 1
\label{Conserv} 
\end{equation}
and has the sense of a quantum ``conservation law" for the dynamics of the nonlinear field. 

\subsection{The backbone map}
In the above we have defined the power $2s$ of the modulus operator $|\hat\psi_n|$ as the power $s$ of the corresponding probability density, i.e., $|\hat\psi_n|^{2s}\equiv (|\hat\psi_n|^{2})^s$. This definition is very nontrivial and in the basis of linearly localized modes leads to 
\begin{equation}
|\hat\psi_n|^{2s} = (\hat\psi_n^{\dag}\hat\psi_n)^s = \left[\sum_{m_1,m_2} \hat a^{\dag}_{m_1} \hat a_{m_2} \phi^*_{n,m_1}\phi_{n,m_2}\right]^s.
\label{2s} 
\end{equation}
Mathematically, it is convenient to consider the power nonlinearity on the right-hand side of Eq.~(\ref{2s}) as a functional map
\begin{equation}
\hat \mathrm{F}_s:\{\phi_{n,m}\}\rightarrow \left[\sum_{m_1,m_2} \hat a^\dag_{m_1} \hat a_{m_2} \phi^*_{n,m_1}\phi_{n,m_2}\right]^s
\label{3s} 
\end{equation}
from the complex vector field $\{\phi_{n,m}\}$ into the ``scalar" field $|\hat\psi_n|^{2s} = (\hat\psi_n^{\dag}\hat\psi_n)^s$. It is noticed that the map in Eq.~(\ref{3s}) is positive definite, and that it contains a self-affine character in it, such that by stretching the basis vectors by a stretch factor $\lambda$ the value of $\hat \mathrm{F}_s$ is renormalized (multiplied by $|\lambda|^{2s}$). We have, accordingly,  
\begin{equation}
\hat \mathrm{F}_s\{\lambda \phi_{n,m}\} = |\lambda|^{2s} \hat \mathrm{F}_s\{\phi_{n,m}\}.
\label{4s} 
\end{equation}
Note that the self-affinity of $\hat \mathrm{F}_s$ is claimed based on the rescaling of the basis vectors $\phi_{n,m}$ and does not involve the commutation properties of the operators $\hat a_m (t)$ and $\hat a_m^{\dag} (t)$. In this regard, the operator form in Eq.~(\ref{3s}) behaves as a $C$-number functional form and adheres to the usual $C^*$-algebra \cite{Arveson}. 

For any nonnegative integer $s$, the power nonlinearity in Eq.~(\ref{2s}) can be expanded in a multinomial series \cite{Stegun}, yielding 
\begin{equation}
|\hat\psi_n|^{2s} = \sum_{\sum {q_{m_1,m_2}} = s}\mathcal{C}_s^{\dots q_{m_1,m_2}}\prod_{m_1, m_2} [\hat \xi_{m_1,m_2}]^{q_{m_1,m_2}},
\label{Multinom} 
\end{equation}
where 
\begin{equation}
\mathcal{C}_s^{\dots q_{m_1,m_2}} = \frac{s!}{\prod_{m_1, m_2} [q_{m_1,m_2}!]}
\label{Coeff} 
\end{equation}
is a multinomial coefficient, the sign $!$ indicates the factorial operation, and we have denoted 
\begin{equation}
\hat \xi_{m_1,m_2} = \hat a^\dag_{m_1} \hat a_{m_2} \phi^*_{n,m_1}\phi_{n,m_2}
\label{Simp} 
\end{equation}
for simplicity. The sum in Eq.~(\ref{Multinom}) is taken over all combinations of nonnegative integer exponents $q_{m_1,m_2}$ such that the sum of all $q_{m_1,m_2}$ is $s$, i.e.,  
\begin{equation}
\sum_{m_1,m_2}{q_{m_1,m_2}} = s.
\label{Sums} 
\end{equation}
An analytic continuation of Eqs.~(\ref{Multinom}) and~(\ref{Coeff}) to noninteger values of $s$ can be obtained by extending the factorial function to the gamma function using $m! = \Gamma (m+1)$ and simultaneously relaxing the condition that the exponents in Eq.~(\ref{Sums}) must be integer. The latter generalization may be achieved {\it iteratively} starting from a situation according to which there is only one such exponent to be accounted for, then gradually increasing the number of the fractional-valued exponents in Eq.~(\ref{Sums}), aiming to assess their overall effect on the final expansion. 

So in the first iteration Eq.~(\ref{Sums}) can only be satisfied if {\it the} fractional exponent that we are looking at (which is the {\it only} fractional exponent in this case) is equal to $s$ exactly (because the sum of the remaining integer-valued exponents cannot add up to a fractional value). Then Eq.~(\ref{Sums}) demands that the sum of the remaining (integer-valued) exponents is zero, and this is an exact result. Assume it is the exponent $q_{i,j}$ which takes the fractional value, i.e., $q_{m_1,m_2} = s$ for some $m_1 = i$ and $m_2 = j$. Then from Eq.~(\ref{Sums}) one infers
\begin{equation}
\sum_{m_1\ne i,m_2\ne j}{q_{m_1,m_2}} = 0.
\label{Dioph} 
\end{equation}
Equation~(\ref{Dioph}) is a Diophantine equation, which is a polynomial equation for which only integer solutions are sought. Because the exponents $q_{m_1,m_2}$ cannot take negative values, the only way Eq.~(\ref{Dioph}) can be satisfied is by setting {\it all} the exponents $q_{m_1,m_2}$ to zero ($m_1 \ne i$, $m_2\ne j$; the exponent for which $m_1 =i$ and $m_2 = j$ is equal to $s$, i.e., $q_{i,j}=s$). It is understood that the polynomial form in Eq.~(\ref{Multinom}) is homogeneous in that the sum of the exponents in each term is always $s$, as Eq.~(\ref{Sums}) shows. On the other hand, the property of the homogeneity implies that any term of the polynomial is in some sense representative of the whole. That means that there is no special reason for which to prefer the very specific setting $m_1 = i, m_2 = j$ against other settings when choosing the fractional-valued exponent, $q_{m_1,m_2}$. The net result is that the condition $q_{i,j} = s$ can be satisfied in a countable number of ways within the range of variation of the parameters $m_1$ and $m_2$. Clearly, all such combinations would equally contribute to the series expansion in Eq.~(\ref{Multinom}). Then to account for these contributions one has to sum over the indexes $m_1$ and $m_2$. Eventually Eq.~(\ref{Multinom}) is simplified to 
\begin{equation}
|\hat\psi_n|^{2s} = \sum_{{m_1,m_2}} [\hat \xi_{m_1,m_2}]^{s}.
\label{Homo} 
\end{equation}
In writing Eq.~(\ref{Homo}) we also used that in the first order   
\begin{equation}
\mathcal{C}_s^{\dots q_{i,j}} = \frac{\Gamma (s+1)}{\Gamma (q_{i,j} + 1)} = \frac{\Gamma (s+1)}{\Gamma (s + 1)} = 1.
\label{First} 
\end{equation}
Substituting $\hat \xi_{m_1,m_2}$ with the aid of Eq.~(\ref{Simp}), from Eq.~(\ref{Homo}) one arrives at 
\begin{equation}
|\hat\psi_n|^{2s} = \sum_{{m_1,m_2}} \hat a^{\dag s}_{m_1} \hat a_{m_2}^s \phi^{* s}_{n,m_1}\phi^s_{n,m_2}.
\label{HF} 
\end{equation}

Let us now consider a more general case where the number of the fractional exponents is at least two or more. This case is complicated by the fact that the sum of two or more fractional numbers may be an integer number; therefore, one cannot separate the fractional and the integer-valued exponents when looking into Eq.~(\ref{Sums}). To this end, we have to refer to our result in Ref. \cite{PRE14} according to which the operators raised to a fractional power that is strictly smaller than 1, i.e., $s < 1$, cannot readily contribute to dynamics due to some sort of topological restrictions in wave number space (to be attributed to the connectedness limitations in the operator space, see below). Then to obtain a nontrivial effect onto field spreading one has to contemplate a nonlinear coupling process among the fractional operators involved, with an idea that such a process would help to generate an effective ``integer" operator first. If the fractional properties are divided between $m+m$ operators ($m$ creation and $m$ annihilation), then to generate one integer creation-annihilation process one needs a pool of $N\sim 2m/s$ fractional operators coming simultaneously into play. Note that $N$ may be a large number, i.e., $N\gg 1$, if the exponent $s$ is small enough. If $p$ is the coupling probability among two arbitrarily chosen waves (for a rarefied lattice gas we may always assume that $p$ is much smaller than 1, i.e., $p\ll 1$), then the coupling probability among $N$ waves would be $p_N \sim p^{N/2}$. It is understood that in a random system this probability will be an exponentially decaying function of $N$, i.e., $p_N \sim \exp [-(N/2)\ln (1/p)]$. That means that the rate of field spreading in a nonlinear random system will be {\it always} dominated by coupling processes in the first order, leading to Eq.~(\ref{HF}) above. 

Our findings so far can be summarized as follows. The reduced model in Eq.~(\ref{HF}) contains all the necessary ingredients to assess the dynamics of field spreading in the original QNLSE model~(\ref{1}). Mathematically, the reduced model derives from an analytic continuation of the multinomial theorem to fractional $s$ values. It uses the idea that in the leading order the coefficients of the multinomial expansion can be obtained by solving a Diophantine equation~(\ref{Sums}) with one fractional index only. 

We note in passing that the reduced model in Eq.~(\ref{HF}) is consistent with the idea that QNLSE~(\ref{1}) is by itself an approximation according to which the nonlinearity $|\hat\psi_n|^{2s}$ occurs as a consequence of the coupling process $|\hat\psi_n|^{2s} = |\hat\psi_n|^{s}\times |\hat\psi_n|^{s}$ in the first order. For $s=1$ (i.e., quadratic power nonlinearity), this approximation is actually quite known in physics \cite{Fishman,Sulem,UFN}. 

Similarly to Eq.~(\ref{3s}) above, the model in Eq.~(\ref{HF}) can be considered as a homogeneous map  
\begin{equation}
\hat \mathrm{F}^\prime_s:\{\phi_{n,m}\}\rightarrow \sum_{m_1,m_2} \hat a^{\dag s}_{m_1} \hat a_{m_2}^s \phi^{* s}_{n,m_1}\phi^s_{n,m_2}
\label{3ss} 
\end{equation}
from the complex vector field $\{\phi_{n,m}\}$ into the operator field in Eq.~(\ref{HF}). In a classical format, we have already encountered this map in Refs. \cite{PRE14,DNC}; where, it was dubbed the ``backbone" map owing to the very peculiar reductions this map offers in the graph space. It was argued that the backbone map preserved (despite the simplifications it carried) the sought dynamical properties of the original QNLSE model, and that it could be considered as representing the algebraic structure of $\hat \mathrm{F}_s$ in the sense of Eq.~(\ref{4s}). Note, in this regard, that the maps $\hat \mathrm{F}_s$ and $\hat \mathrm{F}^\prime_s$ are both {\it self-affine}, obeying the same renormalization rule as of Eq.~(\ref{4s}). That means that a backbone-reduced dynamical model would be characterized by the {\it same} scaling behavior of fluctuating observable quantities, and will lead to the {\it same} scaling laws for transport, as the original model in Eq.~(\ref{1}).    

In view of the above, our further analysis will be based on the backbone-reduced QNLSE, which is obtained by replacing the original functional map $\hat \mathrm{F}_s$ by the backbone map $\hat \mathrm{F}^\prime_s$ for $s < 1$. Note that $\hat \mathrm{F}_s$ coincides with its backbone in the limit $s\rightarrow 1$. This property illustrates the particularity of the quadratic power case {versus} arbitrary power nonlinearity and has been already discussed in Ref. \cite{PRE14}. 

\subsection{Backbone-reduced dynamical model}
Multiplying both sides of the backbone-reduced QNLSE by $\phi^*_{n,k}$ and summing over $n$ with the aid of the orthonormality condition in Eq.~(\ref{Orton}), after simple algebra one obtains the following dynamical equations for the amplitudes $\hat a_{k} (t)$ in the tight-binding approximation:    
\begin{equation}
i\dot{\hat a}_k - \omega_k \hat a_k = \beta \sum_{m_1, m_2, m_3} V_{k, m_1, m_2, m_3} \hat a^{\dag s}_{m_1} \hat a_{m_2}^s \hat a_{m_3},
\label{4s+} 
\end{equation}
where $\omega_k$ is an eigenfrequency of the linear problem, the label $k=0 ,\pm 1,\pm 2, \dots$ takes integer values, the coefficients $V_{k, m_1, m_2, m_3}$ characterize the overlap structure of the nonlinear field and are given by
\begin{equation}
V_{k, m_1, m_2, m_3} = \sum_{n} \phi^*_{n,k}\phi^{*s}_{n,m_1}\phi_{n,m_2}^s\phi_{n,m_3},
\label{5s+} 
\end{equation}
and we have used dot to denote time differentiation. Equations~(\ref{4s+}) correspond to a system of coupled nonlinear oscillators with the Hamiltonian 
\begin{equation}
\hat H = \hat H_{0} + \hat H_{\rm int}, \ \ \ \hat H_0 = \sum_k \omega_k \hat a^{\dag}_k \hat a_k,
\label{6} 
\end{equation}
\begin{equation}
\hat H_{\rm int} = \frac{\beta}{1+s} \sum_{k, m_1, m_2, m_3} V_{k, m_1, m_2, m_3} \hat a^\dag_k \hat a^{\dag s}_{m_1} \hat a_{m_2}^s \hat a_{m_3}.
\label{6+} 
\end{equation}
In the above $\hat H_{0}$ is the Hamiltonian of noninteracting harmonic oscillators, and $\hat H_{\rm int}$ is the interaction Hamiltonian. Note that $\hat H_{\rm int}$ includes self-ineractions through the diagonal elements $V_{k, k, k, k}$. Each nonlinear oscillator with the Hamiltonian   
\begin{equation}
\hat h_{k} = \omega_k \hat a^\dag_k \hat a_k + \frac{\beta}{1+s} V_{k, k, k, k} \hat a^\dag_k \hat a^{\dag s}_{k} \hat a_{k}^s \hat a_{k}
\label{6+h+s} 
\end{equation}
and the equation of motion 
\begin{equation}
i\dot{\hat a}_k - \omega_k \hat a_k - \beta V_{k, k, k, k} \hat a^{\dag s}_{k} \hat a^s_{k} \hat a_{k} = 0
\label{eq+s} 
\end{equation}
represents one nonlinear eigenstate in the system, identified by its wave number $k$, unperturbed frequency $\omega_k$, and nonlinear frequency shift $\Delta \omega_{k} = \beta V_{k, k, k, k} \hat a^{\dag s}_{k} \hat a^s_{k}$. Nondiagonal elements $V_{k, m_1, m_2, m_3}$ characterize couplings between each four eigenstates with the wave numbers $k$, $m_1$, $m_2$, and $m_3$. Resonances occur between the eigenfrequencies $\omega_k$ and the frequencies posed by the nonlinear interaction terms. Then according to Eq.~(\ref{6+}) the resonance condition is 
\begin{equation}
\omega_k = -\omega_{m_1} + \omega_{m_2} + \omega_{m_3}.
\label{Res} 
\end{equation}
Conditions for a nonlinear resonance are obtained by accounting for the nonlinear frequency shift $\Delta \omega_{j}$, where $j=k, m_1, m_2, m_3$ is a resonance wave number. 

When the resonances happen to overlap, a phase trajectory may occasionally switch from one resonance to another, and this will introduce a random ingredient to dynamics in accordance with Chirikov's overlap condition \cite{Sagdeev,ZaslavskyUFN}. Then a nonlinear field may naturally spread along the wave number space via a stochastic process favoring random jumps between the resonances. In this paradigm, an unlimited spreading occurs when the system of overlapping resonances enables a connected escape path to infinity \cite{EPL,PRE14,Chapter}. In general, this path may have a complex organization and be strongly shaped \cite{Chapter,PRE96}. 

\section{Mapping space, Cayley trees and stripes}

Equations~(\ref{4s+}) constitute ever ramifying chains of coupled nonlinear oscillators, with the interaction terms defined by the backbone nonlinearity in Eq.~(\ref{3ss}). Mathematically, it is convenient to represent these chains as an infinite graph by mapping it on a Cayley forest \cite{Schroeder} as follows. 

For each nonlinear eigenstate with the Hamiltonian in Eq.~(\ref{6+h+s}) and equation of motion~(\ref{eq+s}) one finds a node in an countably dimensional mapping space to which a position coordinate $k$ and the associated eigenfrequency, $\omega_k$, are assigned. By {\it countably dimensional} mapping space we mean an Euclidean metric space such that its embedding (integer) dimension is given by the arithmetic number of the different oscillators in Eq.~(\ref{4s+}). Naturally we assume this number to be countable. If we introduce $M = \max \{m_1, m_2, m_3\}$, then the embedding dimension is $d = 2M +1$, provided just that the wave number space is isotropic, i.e., $-M \leq m_1, m_2, m_3 \leq M$. 

Furthermore, the different eigenstates may communicate to each other by exchanging a wave process, and we represent these exchanges by the bonds of the graph connecting the different nodes. These bonds are of two types (see Fig.~1). One type is associated with the ``integer" creation-annihilation operators {\it not} bearing the power mark $s$. Then we represent such operators as simple (connected) bonds; the bond representing the operator $\hat a_{m_3}$ and connecting the node $k$ to $k_3$ in Fig.~1 is an example of this type. The second type is associated with the operators raised to the fractional power $s < 1$; this applies to e.g., the operators $\hat a^{\dag s}_{m_1}$ and $\hat a_{m_2}^s$ in Fig.~1. Following Ref. \cite{PRE14}, we represent such bonds as Cantor sets with the Hausdorff (fractal) dimension $s$. So the Cantor sets in Fig.~1 are the bonds connecting e.g., the node $k_1$ to $k$ and the node $k$ to $k_2$. The wave numbers $m_1$, $m_2$, and $m_3$ identifying the operators $\hat a^{\dag s}_{m_1}$, $\hat a_{m_2}^s$, and $\hat a_{m_3}$ are such that the resonance condition in Eq.~(\ref{Res}) is observed. By examining the structure of Eq.~(\ref{4s+}) one sees that at each step of the communication process there will be exactly $z=3$ bonds (whatever type they are) coming into play: one that we consider ingoing corresponds to the creation operator $\hat a^{\dag s}_{m_1}$, and the other two, the outgoing bonds, correspond to the annihilation operators $\hat a_{m_2}^s$ and $\hat a_{m_3}$, respectively. This gives rise to the characteristic structure of a Cayley tree with the coordination number $z=3$.   

If $s=1$, then the Cantor sets in Fig.~1 become ordinary (simple) bonds, leading to a simplified situation schematically shown in Fig.~2.

\begin{figure}
\includegraphics[width=0.55\textwidth]{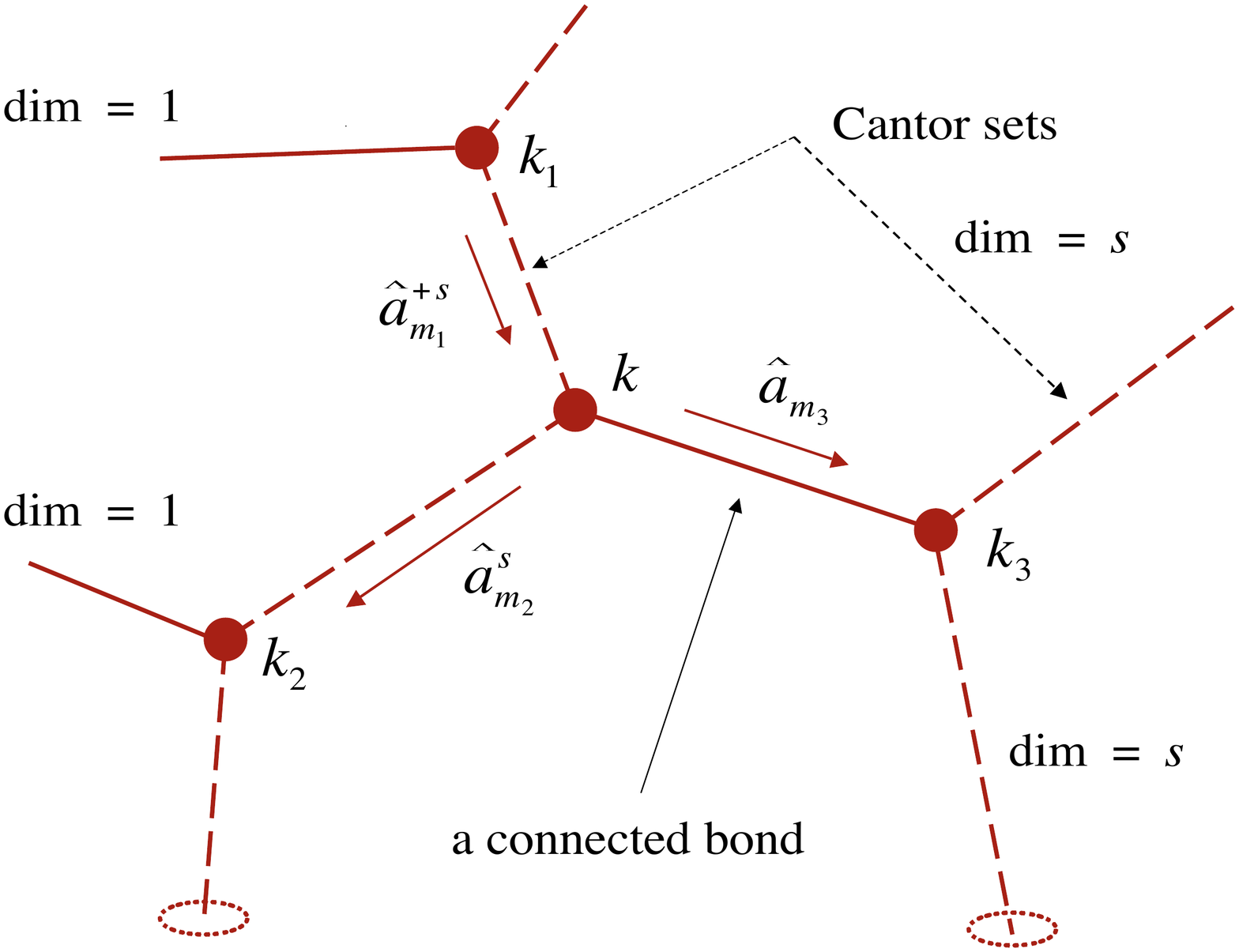}
\caption{\label{} Mapping the system of dynamical equations~(\ref{4s+}) on a Cayley tree. The Cantor sets represent the operators bearing the power label $s$ for $0 < s < 1$ and are shown schematically by the dashed lines. The nodes of the graph are labeled by a position coordinate $k$ and are shown as thick round circles.}
\end{figure}
\begin{figure}
\includegraphics[width=0.55\textwidth]{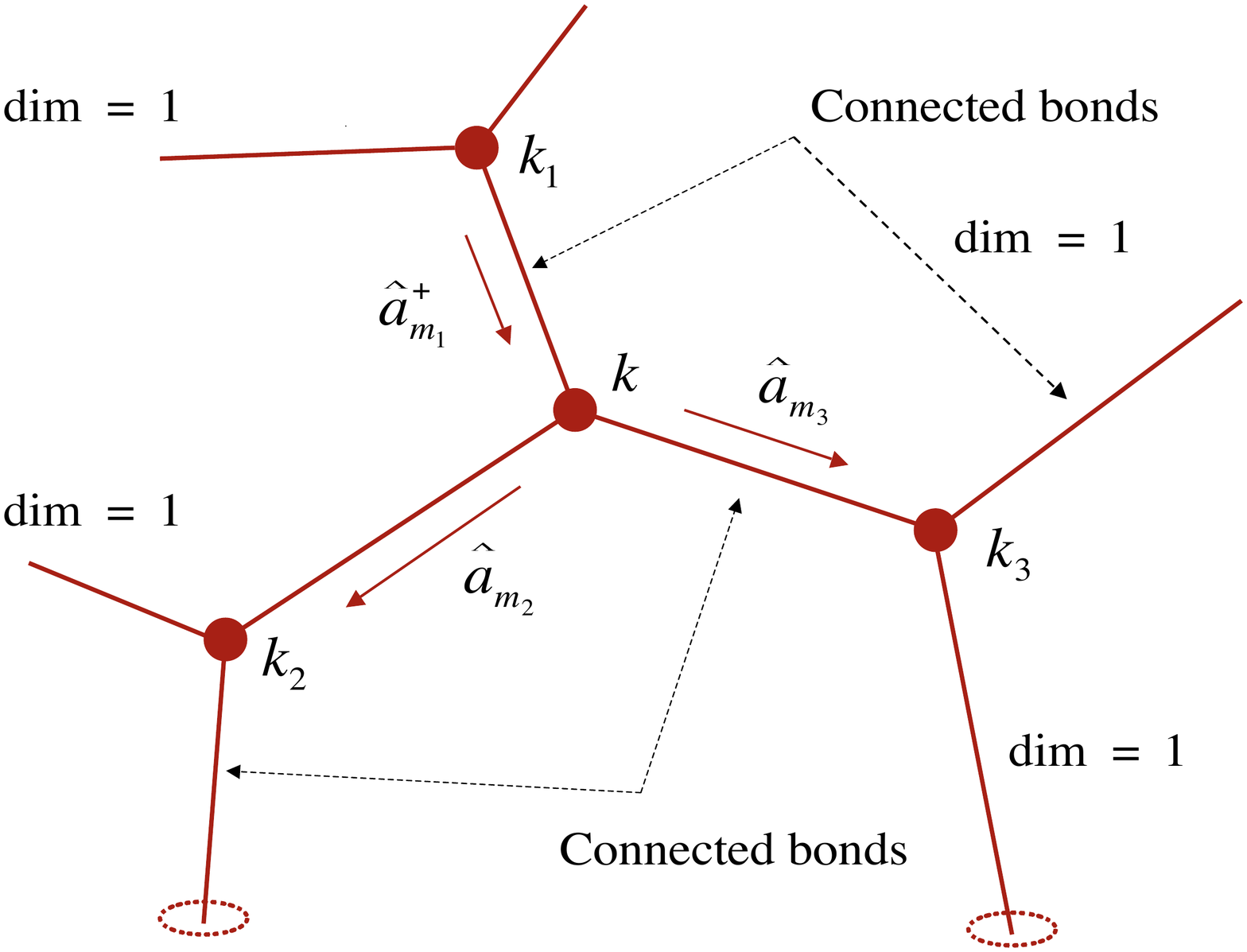}
\caption{\label{} Same situation as in Fig.~1 above, but for $s = 1$. The Cantor sets have become ordinary (simple) bonds (shown by the solid lines).}
\end{figure}

Generally speaking, there exists a number of different connections $(k_1, k_2, k_3)$ to which a given eigenstate with the wave number $k$ can communicate by exchanging a wave process. Indeed, if $M = \max \{m_1, m_2, m_3\}$, then in an isotropic space there exist as many as $2M$ possibilities choosing the wave number $k_1$ (that is, $2M+1$ initial choices minus the choice for the node $k$, already made), consequently $2M-1$ possibilities choosing the wave number $k_2\ne k_1$, and finally $2M-2$ possibilities choosing the wave number $k_3\ne k_2\ne k_1$. Eventually the different arrangements of three different wave numbers would add up to as many as $2M(2M-1)(2M-2)/3!$ combinations, which is none other than the familiar binomial coefficient    
\begin{equation}
\mathcal{C}_{2M}^3 = (2M)!/3!(2M-3)! \sim 4M^3/3.
\label{Binom3} 
\end{equation} 
It is understood that each such combination of indices defines a structural element for a {Cayley tree} originating from the node $k$. For $M\gg 1$, there will be all in all a {\it forest} of incipient Cayley trees, which we arguably call the {Cayley forest} \cite{Schroeder}. Note that the different trees in the forest may occasionally share some nodes or branches and in that case are {\it not} non-intersecting. Because of this, there may occur loops and other ``unpleasant" structural elements that can significantly complicate the analysis of Eq.~(\ref{4s+}). In that regard, it was our suggestion in Refs. \cite{EPL,PRE14} that the system of dynamical equations in Eq.~(\ref{4s+}) could be simplified by assuming that the breakdown of Anderson localization (and the associated transport of the wave field to infinitely long spatial scales) need not occur through {\it all} eigenstates of the nonlinear Schr\"odinger lattice; but, rather, through only a ``critical" connected part of these, such that this part by itself permits a connected escape path to infinity. It has been discussed that this critical part lay on one single Cayley tree and had the topology of a fractal cluster at percolation. Then the outward diffusion of waves from the localization domain could be thought of as a random-walk process along this cluster, leading to the dispersion \cite{EPL}
\begin{equation}
\langle (\Delta n)^2 (t)\rangle \sim t^{1/3}
\label{Binom3} 
\end{equation} 
for $t\rightarrow+\infty$. Also it has been discussed that the reduction of Eq.~(\ref{4s+}) to critical percolation on a tree was only possible for the classical waves and only for $s=1$, and that for $s<1$ the classical transport was hampered by multiple discontinuities due to the Cantor sets in the operator space. Indeed the Cantor sets being discontinuous on all spatial scales implied they could not transmit a classical wave, so a wave initiated at the node $k$, say, could not propagate by more than one step along the tree and ended up at the node $k_3$ (see Fig.~1).   

In this work, we focus on the {\it quantum} transport case and the particularities this case may have with respect to the mechanisms of the transport. {\it We disregard the idea that quantum transport of the wave field in Eq.~(\ref{4s+}) may occur through a critical percolation on a fractal object.} One reason for this is that a quantum wave function would naturally {\it tunnel} between states; then it will be smeared among more states around that are not necessarily restricted to a fractal distribution. For the same reason we would also argue that a quantum wave function could {not} be reduced to one single tree either and that a quantum diffusion would go simultaneously along {\it all} trees defined by Eq.~(\ref{4s+}). The latter property implies that such a transport is {\it not} thresholded \cite{PRE17}, contrary to its classical counterpart. As a consequence, we find that a quantum transport to long distances is possible for all $s \geq 1/2$ and is {\it not} restricted to the quadratic power nonlinearity with $s=1$, at contrast to the classical transport case as of Refs. \cite{EPL,PRE14}.            

The mechanism of quantum spreading, which we discuss, is based on the understanding that the Cayley trees defined by Eq.~(\ref{4s+}) have exactly three ($z=3$) bonds at each their node, and that for $s < 1$ two and only two out of the three such bonds will be Cantor sets which cannot transmit a wave in the classical format. The Cantor sets occur in the mapping space because they represent the operators $\hat a^{\dag s}_{m_1}$ and $\hat a_{m_2}^s$. The latter operators are the usual bosonic operators $\hat a^{\dag}_{m_1}$ and $\hat a_{m_2}$ raised to a fractional power $0 < s < 1$. Arguably we consider the power $s$ as a signature that there is an internal {\it self-interference} taking place for the otherwise ``complete" wave processes $\hat a^{\dag}_{m_1}$ and $\hat a_{m_2}$. Back to Fig.~1 above, the self-interference occurs along the bonds connecting the node $k_1$ to $k$ and the node $k$ to $k_2$. Because $s < 1$, this self-interference is destructive, i.e., the wave tends to cancel itself. The result of this self-cancellation is an ``incomplete" wave process that we arguably call an ``$s$-wave." We consider the $s$-waves as a wave process that has survived the self-cancellation on a fractal object with the Hausdorff measure $s$. It is in this sense that we represented the bonds corresponding to the processes $\hat a^{\dag s}_{m_1}$ and $\hat a_{m_2}^s$ by Cantor sets with the fractal dimension $s < 1$.

\begin{figure}
\includegraphics[width=0.55\textwidth]{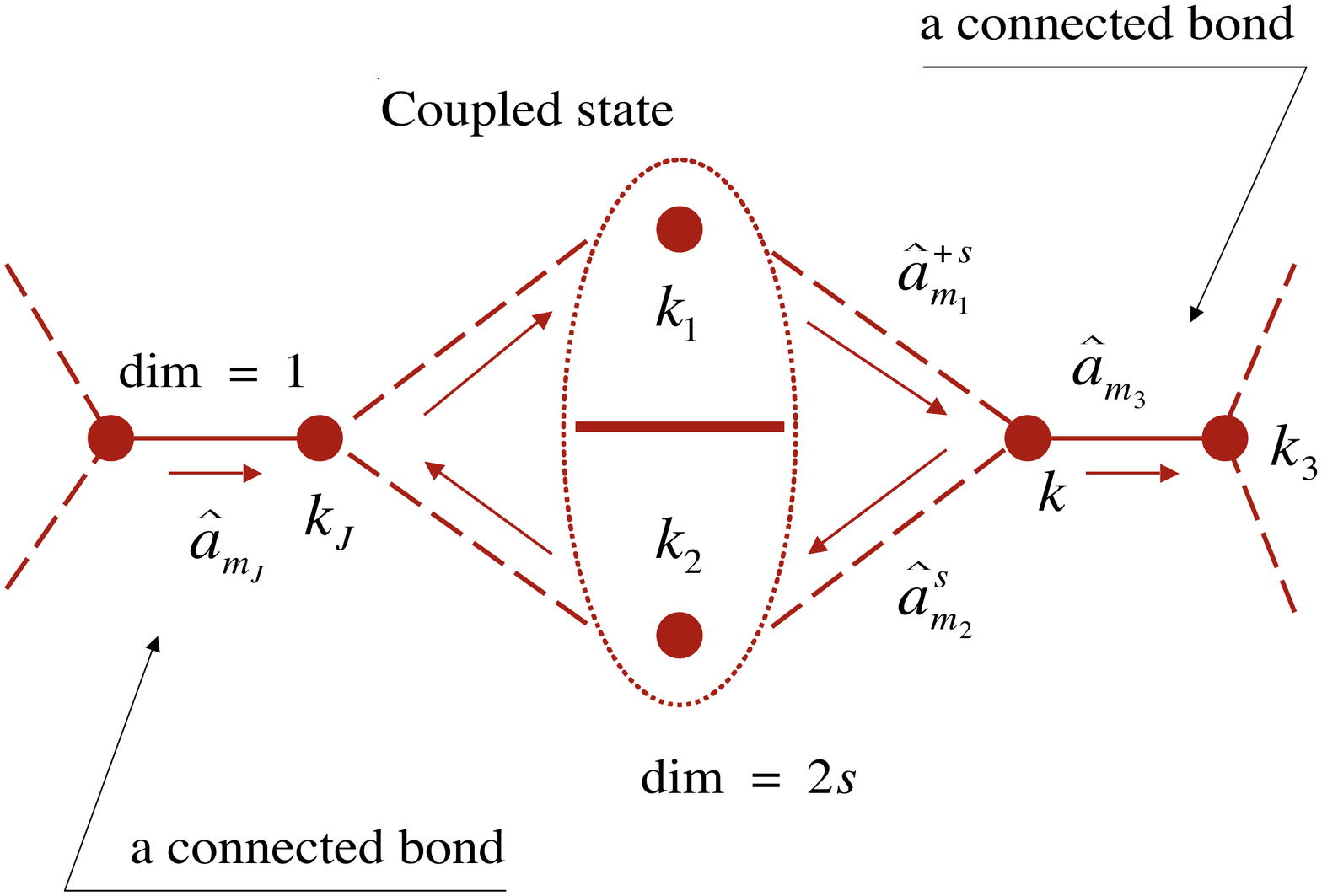}
\caption{\label{} A stripe in the mapping space (schematic illustration). The nodes with the position labels $k_1$ and $k_2$ are identified through quantum smearing of the wave function (shown as an oval encompassing these two nodes). If $s\geq 1/2$, then the coupled state $\hat a^{\dag s}_{m_1}\otimes\hat a_{-m_1}^s$ can accommodate at least one ``complete" wave process, and this is shown as a thick solid line within the oval.}
\end{figure}

In quantum dynamics, the two $s$-waves may tunnel under the topological barriers associated with the Cantor sets$-$a process strictly forbidden in a classical setting. Once the quantum $s$-waves ``meet" under the barrier, they may constructively interfere to produce a nonlinear coupled state which we shall denote by $\hat a^{\dag s}_{m_1}\otimes\hat a_{m_2}^s$. By solving the wave equation for two counter-propagating $s$-waves one may immediately become convinced that the cross-interference is most efficient, if $m_1 = - m_2$, i.e., the momenta of the $s$-waves must be oppositely directed. Let us assume that the constructive interference is ``broad" in that the eigenstates with position coordinates $k_1$ and $k_2$ both lie within a half-width of the same wave function characterizing the coupled state (see Fig.~3). Then these eigenstates cannot be distinguished by the observer and would appear through interactions as one effective, single state. But if the eigenstates at $k_1$ and $k_2$ are inseparable through quantum dynamics, then it is immediate that in the mapping space they must be {\it identified} and hence we must {\it glue} \cite{Nash} them together, so that these nodes become just one single node via the quantum identity relation $k_1 \cong k_2$. 
  
The effect this identity relation has on topology of the mapping space is that it locally affects the {\it connectivity} of this space, giving rise to a theoretical possibility that two otherwise disconnected nodes are connected to each other via a coupled state of two $s$-waves. Naturally one may try to represent this new connection as a bond in the mapping space (shown as thick horizontal line inside the oval in Fig.~3), which in this way of thinking is a substitute for the coupled state $\hat a^{\dag s}_{m_1}\otimes\hat a_{-m_1}^s$ enabling this connection. One sees that such a bond (if it exists) would be characterized by the Hausdorff measure $2s$. In fact, if $s$ is the Hausdorff dimension of the everywhere discontinuous bonds representing the single operators $\hat a^{\dag s}_{m_1}$ and $\hat a_{m_2}^s$, i.e., 
\begin{equation}
\dim [\hat a_{m_1}^{\dag s}] = \dim [\hat a^{s}_{m_2}] = s < 1,\ \ \ \forall m_1, m_2,
\label{Dim-0} 
\end{equation}
then a bond representing a coupled state like $\hat a^{\dag s}_{m_j}\otimes \hat a_{-m_j}^s$ must have the Hausdorff dimension $2s$ in accordance with  
\begin{equation}
\dim [\hat a_{m_j}^{\dag s}\otimes \hat a^{s}_{-m_j}] = \dim [\hat a_{m_j}^{\dag s}] + \dim [\hat a^{s}_{-m_j}] = 2s.
\label{Dim} 
\end{equation}
Equation~(\ref{Dim}) is a manifestation of the general property that the Hausdorff dimension of the direct product of two fractal sets is sum of the respective Hausdorff dimensions of these sets \cite{Isi,PRE96}. Note that the operation ``$\dim$" is defined in the {mapping} space and must not be confused with the ``dimension" of the corresponding Hermitian matrix operator in wave number space. 

Having gone through all this detailed discussion of the subtleties of the connectivity in the mapping space, we may now argue based on Eq.~(\ref{Dim}) that the coupled state $\hat a^{\dag s}_{m_j}\otimes \hat a_{-m_j}^s$ thought of as a quasiparticle state may accommodate at least one ``complete" wave process, if
\begin{equation}
\dim [\hat a_{m_j}^{\dag s}\otimes \hat a^{s}_{-m_j}] = 2s \geq 1,
\label{One-H} 
\end{equation}
i.e., if $s \geq 1/2$. Once a complete wave is there, it may naturally propagate between the nodes $k_1 \cong k_2$ and $k$, and for symmetry reasons also between the nodes $k_J$ and $k_1 \cong k_2$ (see Fig.~3). Here by $k_J$ we mean a ``mirror" node which mediates the response of the coupled state $k_1 \cong k_2$ to an incoming ordinary wave. The end result is that the ``gap" between the otherwise disconnected nodes $k_J$ and $k$ has been filled out through the occurrence of an intermediate coupled state and is now available for quantum transport. 

Our next point here concerns the fact that the coupling process $\hat a_{m_j}^{\dag s}\otimes \hat a^{s}_{-m_j}$ only uses that the momenta of the respective $s$-waves are oppositely directed, but it does not involve an explicit dependence on the position coordinate in the mapping space. If one represents a complete wave processes resulting from the $\hat a_{m_j}^{\dag s} \otimes \hat a^{s}_{-m_j}$ coupling as a connected bond in the mapping space, then based on the dynamical Eq.~(\ref{4s+}) one may construct a countable set of ever continued chains of bonds stretching to arbitrarily long scales along which the nonlinear field may propagate to infinity. Such chains would alternate the creation-annihilation processes defined by the free operators $\hat a_{m_3}$ and other operators alike, with the intermediate quasiparticle states contained in $\hat a_{m_j}^{\dag s} \otimes \hat a^{s}_{-m_j}$ for $m_j = 0,\pm 1,\pm 2, \dots$ We dub such chains of processes ``stripes" whose DNA \cite{DNA} is schematically illustrated in Fig.~3. 

Let us summarize. An important element to quantum transport in disordered nonlinear Schr\"odinger lattices with subquadratic power nonlinearity ($s<1$) is the occurrence of a stripy ordering in the mapping space. The stripes are ever continued (DNA-like) chains of alternating ordinary bonds, on the one hand, and the coupled states of quantum $s$-waves, on the other hand. The stripes form a {\it runway} along which a quantum wave field may propagate to long distances through a disordered structure as a result of the nonlinear interactions behind. Those interactions are captured in a synthetic form by the backbone-reduced dynamical model in Eq.~(\ref{4s+}). Through self-affinity the latter model is characterized by the same scaling properties as the original QNLSE model~(\ref{1}). The stripes occur, because the quasiparticle states resulting from the pairing processes $\hat a_{m_j}^{\dag s} \otimes \hat a^{s}_{-m_j}$ overlap with the propagating wave processes coming with the ordinary creation-annihilation operators. The pairing process $\hat a_{m_j}^{\dag s} \otimes \hat a^{s}_{-m_j}$ is only possible in the quantum domain in that it requires that the corresponding partial $s$-waves tunnel under the topological barriers associated with the (everywhere discontinuous) Cantor sets. 

The occurrence of stripes implies that a nonlinear quantum field cannot be Anderson localized, if $s \geq 1/2$. This quantum result is in marked contrast with the corresponding classical result \cite{EPL,PRE14} according to which a classical nonlinear field is Anderson localized for all $s < 1$ similarly to the linear field. Moreover, the quantum transport is {\it not} thresholded (no onset strength of nonlinear interaction permitting this)$-$contrary to its classical counterpart. The absence of onset strength is again explained by quantum tunneling processes destroying the otherwise thresholded conditions for the critical percolation of a classical wave field on a Cayley tree \cite{EPL}.  

In what follows, we consider the quantum transport along the stripes as a dominant mechanism for the destruction of Anderson localization in quantum nonlinear Schr\"odinger lattices with $s\geq 1/2$. Note, in this regard, that by looking into a stripy ordering we may appreciably simplify the analysis and restrict ourselves just to one-dimensional transport models.

\section{The Lennard-Jones potential and subdiffusive spreading law}

It is understood that the excitation of each eigenstate is none other than the spreading of the wave field in wave number space \cite{PS,EPL}. If the field is spread over $\Delta n \gg 1$ states, then the conservation of the total probability 
\begin{equation}
\sum_n \hat\psi_n^{\dag}\hat\psi_n \simeq \int |\hat\psi_n|^2 d\Delta n = \hat 1
\label{TP} 
\end{equation}
would imply that $|\hat\psi_n|^2 \simeq \hat 1/\Delta n$. In the basis of linear localized modes, the evolution of the operators ${\hat a}_m = {\hat a}_m (t)$ is due to their nonlinear coupling, i.e., 
\begin{equation}
\dot{\hat a}_m \sim \beta \hat a_{m_1}^{\dag s} \hat a^{s}_{m_2} \hat a_{m_3}.
\label{NLC} 
\end{equation}
The rate of excitation of the newly involved modes at the site $m$ is obtained as $R_m \sim |\dot{\hat a}_m|^2$ and will be proportional to the power $2s +1$ of the probability density. The system-average mode excitation rate is written as $R \sim |\dot {\hat \psi_n}|^2 \sim |\hat \psi_n|^{2(2s+1)}$. Taking the conservation of the total probability into account, one arrives at 
\begin{equation}
R\sim \beta^2/ (\Delta n)^{2s + 1}.
\label{Rate} 
\end{equation}
On the other hand, the number of the newly excited modes per unit time is $d\Delta n /dt$, making it possible to assess $d \Delta n / dt \sim \beta^2/ (\Delta n)^{2s + 1}$. The latter condition is different from the corresponding condition used in Refs. \cite{PS,PRE14} in that we do not assume that the spreading of the wave field is of the diffusive type; nor do we involve any sort of random-phase approximation justifying such an assumption. Indeed, in quantum dynamics, the notion of chaos loses its classical meaning \cite{Casati}. Therefore, the time derivative $d/dt$ is applied to $\Delta n$ itself$-$as dictated by Fermi's golden rule \cite{Golden}$-$and not to the square of $\Delta n$, as of Refs. \cite{PS,EPL,PRE14}, leading to a different law of spreading. Before we proceed, we abolish the similarity sign in $d \Delta n / dt \sim \beta^2/ (\Delta n)^{2s + 1}$ in favor of the equation $d \Delta n / dt = A / (\Delta n)^{2s + 1}$ in which $A\propto \beta^2$ has absorbed all numerical coefficients behind. Integrating over time, one gets $(\Delta n)^{2s + 2} = (2s + 2) A t$, from which a subdiffusive law of spreading     
\begin{equation}
(\Delta n)^2 = [(2s + 2) A]^{1/(s+1)}\,t^{1/(s+1)}
\label{Sub} 
\end{equation}
can be deduced. For $s\rightarrow 1$, the behavior on the right-hand side of Eq.~(\ref{Sub}) is square-root-like (as a nickname, we would call this ``half-diffusion"). The half-diffusion $(\Delta n)^2 \propto t^{1/2}$ corresponds to a faster spreading process as compared to Eq.~(\ref{Binom3}). The explanation lies in the fact that the quantum specific phenomena, such as tunneling between states, etc., naturally enhance the transport above the classically expected values.  

Let us now assess the dynamics of field spreading from the perspective of the second-order time derivative. For this, differentiate the equation $d \Delta n / dt = A / (\Delta n)^{2s + 1}$ with respect to time, then eliminate on the right-hand side the first derivative $d\Delta n /dt$ using the same equation. The end result is $d^2 \Delta n /dt^2 = -(2s + 1)A^2 / (\Delta n)^{4s + 3}$. Rewriting the power-law dependence on the right-hand side such that it takes the form of a ``gradient" in the $\Delta n$ direction, one gets
\begin{equation}
\frac{d^2}{dt^2}\Delta n = - \frac{d}{d \Delta n} \left[- \frac{A^2 / 2}{(\Delta n)^{4s + 2}}\right].
\label{Grad} 
\end{equation}
So, if $\Delta n$ is a position coordinate in wave number space, as in fact it is, then Eq.~(\ref{Grad}) is none other than the Newtonian equation of motion in the potential field 
\begin{equation}
W (\Delta n) = - \frac{A^2 / 2}{(\Delta n)^{4s + 2}}.
\label{Poten} 
\end{equation}
For $s\rightarrow 1$, the potential function in Eq.~(\ref{Poten}) takes the form $W(\Delta n) = -(A^2 / 2)/ (\Delta n)^6$, which is immediately recognized as the attractive part of the celebrated Lennard-Jones potential \cite{Lennard}. The latter potential finds outstanding applications \cite{Rapa} in molecular dynamics and quantum chemistry. As a result of the attracting dynamics in Eq.~(\ref{Grad}), the newly excited modes will tend to form clusters$-$``molecules"$-$in wave number space; where, they will be effectively trapped due to their nonlinear coupling \cite{Iomin}. The comprehension of the attractive ``forces" between the components of the wave field explains the deviation from the normal diffusion in the nonlinear Schr\"odinger dynamics. Indeed the transport is subdiffusive, i.e., $(\Delta n)^2 \sim t^{1/(s+1)}$, and not $\sim t$ as in the Brownian transport case, owing to the binding effect of the potential field of the Lennard-Jones type. We shall illustrate this property shortly.   

Multiplying both sides of Eq.~(\ref{Grad}) by the ``velocity", $d \Delta n /dt$, and integrating the ensuing differential equation with respect to time, after simple algebra one obtains
\begin{equation}
\frac{1}{2}\left[\frac{d}{dt} \Delta n\right]^2 - \frac{A^2 / 2}{(\Delta n)^{4s + 2}} = \Delta E,
\label{Ener} 
\end{equation}
where the first term on the left-hand side has the sense of the kinetic energy of a ``particle" of unit mass moving along the $\Delta n$ coordinate, and the second term is its potential energy. It is shown using the equation $d \Delta n / dt = A / (\Delta n)^{2s + 1}$ that the kinetic energy in Eq.~(\ref{Ener}) compensates for the potential energy {\it exactly}, that is, the full energy in Eq.~(\ref{Ener}) is actually zero, $\Delta E = 0$. More so, both the negative potential energy $W (\Delta n) \sim - A^2 / (\Delta n)^{4s + 2}$ and the positive kinetic energy $\frac{1}{2}(d \Delta n /dt)^2 \sim A^2 / (\Delta n)^{2(2s + 1)}$ vanish while spreading. Both will decay as the $(4s + 2)$-th power of the number of states, and the ratio between them will {\it not} depend on the width of the field distribution. 

The full energy being equal to zero implies that the ``particle" in Eq.~(\ref{Ener}) is sitting on the separatrix $\Delta E = 0$. Based on the analysis of Ref. \cite{PRE00} we may argue that the separatrix $\Delta E = 0$ contains a connected escape path to infinity; hence it allows for an unlimited spreading of the wave field regardless of the strength of the nonlinearity. More so, as the particle propagates outward, its motion becomes intrinsically unstable (sensitive to fluctuations). This is because both the potential and the kinetic energies vanish for $\Delta n \rightarrow +\infty$, so very tiny perturbations due to for instance random noise, zero point fluctuations, quantum tunneling, etc. may drastically change the type of phase space trajectory. The result generally holds for the dynamics near separatrices in large systems \cite{Sagdeev,ZaslavskyUFN,ChV}. To this end, the fact that a given mode does or does not belong to a cluster of modes becomes essentially a matter of the probability. 

To assess the probabilistic aspects of field spreading, let us assume that the fluctuation background is characterized by the effective thermodynamic ``temperature", $T$. So, the value of $T$ weighs all occasional perturbations to dynamics that might be influential near the separatrix. Then the probability for a given mode to quit the cluster after it has traveled $\Delta n$ sites on it can be written as the Boltzmann factor $p (\Delta n) = \exp [W (\Delta n) / T]$, where we have set the Boltzmann constant to 1 for simplicity. Substituting $W (\Delta n)$ from the Lennard-Jones potential in Eq.~(\ref{Poten}), one gets 
\begin{equation}
p (\Delta n) = \exp [- A^2 / 2 T (\Delta n)^{4s + 2}].
\label{Escape} 
\end{equation}
Taylor expanding the exponential function for $\Delta n \gg 1$, we find $p (\Delta n) \simeq 1 - A^2 / 2 T (\Delta n)^{4s + 2}$. The probability to remain (``survive") on the cluster after $\Delta n$ space steps is $p^{\prime} = 1-p$, yielding 
\begin{equation}
p^{\prime} (\Delta n) \simeq A^2 / 2 T (\Delta n)^{4s + 2}.
\label{EscapePr} 
\end{equation}
Eliminating $\Delta n$ with the aid of Eq.~(\ref{Sub}), one obtains the probability to survive on the cluster for $t$ time steps, i.e.,  
\begin{equation}
p^{\prime} (t) \propto t^{-(2s+1) / (s+1)},
\label{Survive} 
\end{equation}
where we have omitted the redundant dimensional coefficient $A^{1/(s+1)} / 2T (2s+2)^{(2s + 1)/(s+1)}$ in front of the scaling factor, $\propto t^{-(2s+1) / (s+1)}$. Statistically, one may interpret the survival probability in Eq.~(\ref{Survive}) as a waiting-time distribution $\chi (t) \propto t^{-(2s+1) / (s+1)}$ reflecting the binding effect of the different clusters in wave number space. Note that $t$ in Eq.~(\ref{Survive}) is a time scale, which has the sense of exit$-$or trapping \cite{Bouchaud}$-$time for the random walks on finite clusters. Note, also, that the integral $\int t \chi (t) dt\sim t^{1/(s+1)}$ diverges for $t\rightarrow+\infty$, implying that the mean waiting time is infinite.

\section{Bifractional kinetic equation}

In this section, we devise a kinetic picture for the asymptotic ($t\rightarrow+\infty$) transport using random walks and the formalism of a fractional-derivative diffusion equation. Indeed the fractional kinetic equations dealing with generalized derivatives in space and time incorporate in a natural, unified way the key features of non-Gaussianity and long-range dependence that often break down the restrictive assumptions of locality and lack of correlation underlying the conventional statistical mechanical paradigm \cite{Klafter,Klafter2004,Sokolov}. From a probabilistic standpoint, fractional kinetics extends Gaussian stochastic processes (i.e., Brownian random walks) by taking into account long-range correlated events in the tail of the probability density function. Mathematically, the use of fractional-derivative equations is advantageous, as it makes it possible to describe transport phenomena in complex systems in much the same way as in simpler systems \cite{Sokolov}.

\subsection{Continuous time random walks in wave number space}

We propose based on the above analysis that the asymptotic transport in the QNLSE model~(\ref{1}) occurs in the form of a random walk process along the stripes and we adopt the standard scheme of continuous time random walks (CTRW) \cite{Bouchaud,CTRW} to describe this process. In this paradigm one assumes that the random walker (a component of the wave field, in our case) jumps in random direction along the coordinate axis and that the jumps are such that there is a distribution of waiting times between consecutive steps of the random motion, on the one hand, and a distribution of the jump lengths in the coordinate space, on the other hand. Then we adopt the following paradigmatic distributions respectively for the waiting times and the jump lengths ($t$ is the waiting time; $\ell$ is the jump length), i.e.,     
\begin{equation}
\chi (t) \propto 1/(1+t/\tau)^{1+\alpha},
\label{WaitingT} 
\end{equation} 
\begin{equation}
\chi (\ell) \propto A_\mu \ell^{-(1+\mu)},
\label{JumpL} 
\end{equation} 
with $0 < \alpha < 1$ and $1 < \mu < 2$ (see Refs. \cite{Klafter,Sokolov}). In the above $\tau$ is the microscopic lattice time and is set to 1 for simplicity, and $A_\mu$ is a normalization constant parameter. Then starting from the model distributions in Eqs.~(\ref{WaitingT}) and~(\ref{JumpL}), and using that the steps occur in random direction along a one-dimensional coordinate axis, one obtains the following kinetic equation for the probability density $f = f(t, \Delta n)$ to find the random walker at time $t$ at the distance $\Delta n$ from the origin \cite{Klafter,Klafter2004}:   
\begin{equation}
\frac{\partial}{\partial t} f (t, \Delta n) =\,_{0}\mathrm{D}_t^{1-\alpha}[\kappa_{\alpha, \mu} \nabla^{\mu} f (t, \Delta n)].
\label{FDEL} 
\end{equation}
Here, $_{0}\mathrm{D}_t^{1-\alpha}$ is the Riemann-Liouville fractional derivative with respect to time, and $\nabla^{\mu}$ is the Riesz fractional derivative and is taken with respect to the position coordinate $\Delta n$ (see Eqs. (A.4) and (A.13) of Ref. \cite{Klafter}). Further $\kappa_{\alpha, \mu}$ is the transport coefficient, which carries the dimension cm$^{\mu}\,\times\,$s$^{-\alpha}$. The Riemann-Liouville derivative is defined by 
\begin{equation}
_{0}\mathrm{D}_t^{1-\alpha}f (t, \Delta n) = \frac{1}{\Gamma (\alpha)}\frac{\partial}{\partial t}\int _{0}^{t} dt^{\prime} \frac{f (t^{\prime}, \Delta n)}{(t - t^{\prime})^{1-\alpha}}
\label{RL} 
\end{equation}
and accounts for the non-Markovian properties of the transport in the limit $t\rightarrow+\infty$, e.g., multi-scale trapping phenomena along the stripes, stickiness, non-ergodicity and other phenomena alike \cite{Report}. Note that the Riemann-Liouville derivative is none other than the ordinary time derivative applied to a convolution of the probability density with a power-law. By setting the lower limit of the time integration to zero we have tacitly assumed that the random walk process is initiated at time $t=0$. Note, in this regard, that the Riemann-Liouville derivative incorporates the initial value problem through the definition of the lower limit of the integration. 

Concerning the Riesz fractional derivative in Eq.~(\ref{FDEL}), its definition \cite{Ch2007,Klafter2004} is such as to achieve a meaningful generalization of the Laplacian operator to systems in which the standard Gaussian central limit theorem \cite{Gnedenko} is invalidated as a result of the heavy-tailed jump length distribution in Eq.~(\ref{JumpL}). In our notation 
\begin{equation}
\nabla^{\mu} f (t, \Delta n) = \frac{1}{\Gamma_{\mu}}\frac{\partial^2}{\partial \Delta n^2} \int_{-\infty}^{+\infty}\frac{f (t, \Delta n^\prime)}{|\Delta n-\Delta n^\prime|^{\mu - 1}} d\Delta n^\prime,
\label{Riesz} 
\end{equation}
where $\Gamma_{\mu} = - 2\cos(\pi\mu/2)\Gamma (2-\mu)$ is a normalization parameter, and $1 < \mu < 2$ is the fractional index of the integro-differentiation [same $\mu$ in Eq.~(\ref{JumpL}) above]. The improper integral on the right-hand side of Eq.~(\ref{Riesz}) is understood as the sum of two Riemann-Liouville integrals, i.e., $\int_{-\infty}^{+\infty} = \int_{-\infty}^{\Delta n} + \int_{\Delta n}^{+\infty}$. The integration in Eq.~(\ref{Riesz}) is performed in infinite limits along the position coordinate $\Delta n$ and in the mapping space corresponds to an improper integration along a stripe. For $\mu\rightarrow 2$, the conditions of the standard Gaussian limit theorem are reinstalled. In that case one naturally uses in Eq.~(\ref{FDEL}) the standard Laplacian operator $\nabla^2 = \partial^2 / \partial (\Delta n)^2$ in place of the Riesz operator, $\nabla^\mu$. For $\mu \rightarrow 1$, the Riesz derivative reduces (through the degeneration of the normalization parameter, $\Gamma_\mu\rightarrow +0$) to the Hilbert transform operator \cite{Klafter2004,Mainardi}     
\begin{equation}
\lim_{\mu\rightarrow 1}\nabla^{\mu} f (t, \Delta n)  = - \frac{1}{\pi} \frac{\partial}{\partial \Delta n} \int_{-\infty}^{+\infty}\frac{f (\Delta n^\prime, t)}{\Delta n-\Delta n^\prime} d\Delta n^\prime.
\label{Hilbert} 
\end{equation} 
A derivation of kinetic Eq.~(\ref{FDEL}) is articulated in e.g., Refs. \cite{Klafter,Ch2007,Jesper,Gonchar} for systems driven by a stochastic noise process. A derivation using transition probabilities in reciprocal space has been obtained in Refs. \cite{Transit,PRE2018}. As is well known, the Riesz operator in Eq.~(\ref{FDEL}) generates L\'evy flights \cite{Klafter,Ch2007}. A L\'evy flight is defined as a motion process driven by an uncorrelated L\'evy ``white" noise \cite{Ch2007}. The defining feature of L\'evy flights is their ability to propagate {\it nonlocally} to long distances by performing instantaneous jumps in ambient space, with a jump-length statistics as of Eq.~(\ref{JumpL}). The Hilbert case in Eq.~(\ref{Hilbert}) is special and mathematically corresponds to the Cauchy flights identified by the limiting value $\mu = 1$. We consider the Cauchy flights as a natural bound on the L\'evy-type processes discussed in this work.  

In a basic theory of L\'evy flights one shows that the L\'evy flight trajectory possesses a fractal dimension characterizing the island structure of clusters of smaller steps, connected by a long step (Ref. \cite{Klafter}, p. 27). This fractal dimension is given by $d_f = \mu$. The L\'evy flight trajectory being a continuous fractal curve implies that its fractal dimension is not smaller than 1, i.e., $d_f \geq 1$. We have, accordingly, $\mu\geq 1$. 

Mathematically, the nonlocality of L\'evy flights is included via the improper integration in Eqs.~(\ref{Riesz}) and~(\ref{Hilbert}). The fact that the convolution is taken with a power-law is consistent with the property of the homogeneity enabling the scaling relation in Eq.~(\ref{4s}) and the introduction of a backbone-reduced dynamical model in Eq.~(\ref{4s+}). A boundary-value problem for L\'evy flights is discussed in e.g., Refs. \cite{Klafter2004,PRE2018,Chechkin2003}.

\subsection{Values of fractional exponents}

In a statistical perspective, Eq.~(\ref{FDEL}) is a starting point to obtain fractional moments of the probability density function. The exact calculation uses the Mellin transformation and the formalism of Fox functions \cite{Klafter,Samko}. Here, we restrict ourselves to a qualitative result contained in the scaling relation $(\Delta n)^{\mu} \propto t^{\alpha}$, from which a pseudo mean-square displacement $[\Delta n]^{2} \propto t^{2\alpha/\mu}$ can be inferred for long times. This is consistent with Eq.~(\ref{Sub}) above, if 
\begin{equation}
2\alpha / \mu = 1 / (s+1).
\label{MuA} 
\end{equation} 
The $\alpha$ value is obtained by comparing the waiting-time distributions in, respectively, Eqs.~(\ref{Survive}) and~(\ref{WaitingT}), the result being 
\begin{equation}
\alpha = s / (s+1).
\label{SV} 
\end{equation} 
This value is {\it self-consistent}. Substituting this in Eq.~(\ref{MuA}), one obtains the corresponding value of the fractional exponent $\mu$ to be 
\begin{equation}
\mu = 2s.
\label{MuS} 
\end{equation} 
The result in Eq.~(\ref{MuS}) is not really surprising. It means that the rate of excitation of the newly excited states and the jump length distribution between clusters of states both belong to the same power-law, with the same drop-off exponent, as a comparison of Eqs.~(\ref{Rate}) and~(\ref{JumpL}) shows. This is to be expected, since the L\'evy flyer jumping between two points in wave number space would excite dynamically a whole cluster of states lying between these points. Since the flight trajectories are continuous curves, as we have assumed they are, then the condition in Eq.~(\ref{MuS}) is readily inferred for $\mu \geq 1$ based on the connectedness arguments. The end result is that the transport to infinite distances occurs in the QNLSE model~(\ref{1}), if $\mu = 2s \geq 1$, i.e., 
\begin{equation}
s \geq 1/2.
\label{End_R} 
\end{equation} 
The latter is consistent with the analysis of Sec. III based on the idea of sub-barrier coupling among the $s$-waves [see Eq.~(\ref{One-H})]. If the transport proceeds as a competition between the waiting-time statistics and L\'evy flights, then the pseudo mean-square displacement scales as $[\Delta n]^{2} \propto t^{1/(s+1)}$. Setting $s=1$, one gets $(\Delta n)^{2} \propto t^{1/2}$ reviving the finding of Ref. \cite{PRE17}. 

More so, if $s=1$, then the index of fractional integro-differentiation in Eq.~(\ref{Riesz}) takes its limiting (integer) value $\mu = 2$. The implication is that L\'evy flights would only occur for subquadratic power nonlinearities, with $s < 1$; whereas in the quadratic power case ($s=1$) the transport goes as a non-Markovian diffusion with a waiting-time distribution in Eq.~(\ref{WaitingT}), where $\alpha = 1/2$. This process is such that it adopts a characteristic jump length in wave number space and corresponds to the conditions of the standard Gaussian central limit theorem \cite{Gnedenko}. This situation has been discussed in our previous publication \cite{PRE17}. Based on this evidence, one is led to conclude that the L\'evy flights are a characteristic of a QNLSE model with the fractional $s$ values that must be strictly smaller than 1, i.e., $s < 1$.

\begin{table}[t]
\begin{center}
\begin{tabular}{p{2.0cm}p{2.2cm}p{2.2cm}p{1.7cm}} \hline \hline
Quantum & $0\leq s < 1/2$ & $1/2 \leq s < 1$  & $s=1$  \\ \hline 
Localization & Yes & No  & No \\
Spreading & No & Yes  & Yes \\
Thresholded & N/A & No  & No \cite{PRE17} \\
Stripes & No & Yes & Yes\footnote{In this limit, the stripes have the fractal dimension $d_f = 2$ (same as the Brownian random walk \cite{Klafter}) and in this sense are {\it not} low-dimensional.} \\ 
L\'evy flights & No & Yes & No\footnote{The jump-length distribution is Gaussian in this case. The implication is that there is a characteristic jump length, rather than a heavy-tailed distribution of these as of Eq.~(\ref{JumpL}).} \\
Trappings & Yes & Yes & Yes \\
Transport & No & Subdiffusive\footnote{Occurs as a competition between long-time trappings with a distribution of waiting/exit times, on the one hand, and L\'evy flights along the stripes, on the other hand.} & Subdiffusive \\
Dispersion\footnote{Pseudo mean-square displacement in case of subdiffusion with L\'evy flights ($1/2 \leq s < 1$); otherwise the usual mean-square displacement $\langle (\Delta n)^2 (t)\rangle$.} & N/A & $t^{1/(s+1)}$  & $t^{1/2}$ \cite{PRE17}\\
\hline\hline
Classical & $0\leq s < 1/2$ & $1/2 \leq s < 1$  & $s=1$  \\ \hline 
Localization & Yes & Yes  & No\footnote{Above a certain critical strength of nonlinear interaction \cite{PS,EPL}.} \cite{PS, EPL} \\
Spreading & No & No  & Yes\footnote{Above a certain critical strength of nonlinear interaction \cite{PS,EPL}.} \cite{PS,EPL} \\
Thresholded & N/A & N/A  & Yes \cite{EPL,PRE14} \\
Stripes & No & No & No \\ 
L\'evy flights & No & No & No \\
Trappings & Yes & Yes & Yes\footnote{Not in chaotic case, see below} \\
Transport & No & No & Subdiffusive \\
Dispersion\footnote{Pseudochaotic \cite{Report} random walks at the threshold of delocalization \cite{EPL,Chapter,PRE14,DNC}.}  & N/A & N/A  & $t^{1/3}$ \cite{EPL,PRE14} \\
Dispersion\footnote{Chaotic diffusion with a range-dependent diffusion coefficient above the delocalization point \cite{EPL,PRE14}.} (chaotic) & N/A & N/A  & $t^{2/5}$ \cite{PS,EPL} \\
\hline \hline
\end{tabular}
\end{center}
\caption{A summary of results and comparison between the quantum and classical situations for the different values of the exponent $s$. One sees that the quantum transport is generally faster than its classical counterpart (for $s=1$) and it also occurs in the parameter range $1/2 \leq s < 1$ for which no classical transport has been found \cite{PRE14,DNC}. N/A means ``not appropriate" (wherever the issue is undefined or badly posed).} \label{tab1}
\label{default}
\end{table}

In Table~I, we summarize our results and also compare the quantum and the classical transport cases with a special focus on similarities and differences between the two transport regimes, such as for instance the occurrence of stripes, L\'evy flights, and a waiting-time statistics. While we do not discuss the classical transport here, an interested reader can easily refer to Refs. \cite{EPL,PRE14,Chapter,DNC} in which further particularities may be found. 

\section{Discussion}

We proceed with a few remarks. Our first point here concerns the use of the CTRW scheme and the fact that the spreading process by L\'evy flights involves an important chaotic ingredient into dynamics. We argue that this ingredient is {\it self-assembling} and occurs {naturally} through the phenomena of quantum tunneling. This is made more precise in the following.  

\subsection{The chaos is self-reinforcing}

The focus here is on Chirikov's overlap parameter $K \simeq \Delta\omega_{\rm NL} / \delta \omega$ \cite{Sagdeev,ZaslavskyUFN,Report} which in our case characterizes the extent to which single resonances in Eq.~(\ref{Res}) can overlap due to their nonlinear broadening. The $K$ value is defined as a ratio between the nonlinear frequency shift $\Delta\omega_{\rm NL} = \beta |\hat\psi_n|^{2s}$ and the typical distance between resonances in wave number space, $\delta\omega$. The latter distance scales with the number of states $\Delta n$ as $\delta \omega \simeq 1 / \Delta n$. The overlap parameter being much larger than one implies that the chaos is strong paving the way for a kinetic description in terms of the probability density function $f = f (t, \Delta n)$. With the aid of Eq.~(\ref{TP}) one finds that $|\hat\psi_n|^2 \simeq \hat 1/\Delta n$ for $\Delta n \gg 1$, leading to $\Delta\omega_{\rm NL} \simeq \beta / (\Delta n)^{s}$. Therefore, $K\simeq \beta (\Delta n)^{1-s}$.

One sees that for $s < 1$ the $K$ value involves a dependence on the number of states. Through quantum tunneling this number is a non-decreasing function of time, i.e., $d\Delta n / dt \geq 0$. So, if the initial condition is such that $K_{t=0}\gg 1$, then the property for the overlap parameter to be large will be preserved while spreading. Moreover, the chaos is self-reinforcing in that  
\begin{equation}
dK / dt \simeq \beta (1-s) (\Delta n)^{-s} d\Delta n / dt \geq 0.
\label{SRI} 
\end{equation} 
In our previous works \cite{EPL,PRE14} we have argued that the nonlinear oscillators at each node $k$ could be found in one of two states$-$either chaotic (``dephased") or regular, and that it was up to the oscillators in the chaotic state to re-emit the waves further, thus favoring a large-scale transport in wave number space. Also we have argued that the probability for an oscillator to be found in a chaotic state was given by the Boltzmann factor 
\begin{equation}
p_{\Delta\omega_{\rm NL}} = \exp (-\delta\omega / \Delta\omega_{\rm NL}) = \exp [-1/\beta (\Delta n)^{1-s}]
\label{B-factor} 
\end{equation}
and that the nonlinear frequency shift $\Delta\omega_{\rm NL}$ had the sense of the effective ``temperature" of the nonlinear interaction. If the parameter $s$ is strictly smaller than 1, then Eq.~(\ref{B-factor}) would imply that $\lim_{t\rightarrow+\infty} p_{\Delta\omega_{\rm NL}}(t) = 1$, because $\Delta n \rightarrow+\infty$ for $t\rightarrow+\infty$ as a result of quantum tunneling. That is, the asymptotic state of coupled nonlinear oscillators in the dynamical system~(\ref{4s+}) is the chaotic state (at no contradiction to the fact that the oscillators may be organized in clusters introducing an exit-time statistics). In view of Eq.~(\ref{SRI}), the chaotic state is an {\it attractor} for the dynamics. 

As the oscillators form stripes through quantum tunneling of counter-propagating $s$-waves, one says the stripes are a channel through which the chaotic motions can propagate to infinitely long distances in wave number space, thus destroying the Anderson localization by nonlinear interactions. In that regard, the assumption that the transport is driven by a L\'evy {\it white} noise finds its justification in the chaotic character of the interactions. We note in passing that the dependence on the $\beta$ value in Eq.~(\ref{B-factor}) is {\it non-perturbative} for $\beta\rightarrow 0$, and that the linear model, with $\beta = 0$, is characterized by the oscillators residing all in the regular state, which does not conduct the waves. For $s=1$ exactly, the onset of long-distance transport is limited to the $\beta$ value, which must be large enough to guarantee a sufficiently broad population of the dephased oscillators conducting the wave processes to infinity \cite{EPL,PRE14}.   
  
\subsection{Self-organization without criticality}

Another point worth noting is that the asymptotic state is characterized by a power-law waiting/exit time distribution with the diverging mean waiting time (see Eqs.~(\ref{Survive}) and~(\ref{WaitingT})) and, simultaneously, by a Pareto-L\'evy distribution of jump lengths as of Eq.~(\ref{JumpL}). More so, both distributions are parametrized by the same exponent $s < 1$, making it possible to express the $\alpha$ value in terms of the fractional index $\mu$ as $\alpha = \mu / (\mu + 2)$. The implication is that the asymptotic transport is such that there is an important coupling between statistical properties in time and in space. These properties are, moreover, {\it not} separable. Often the phenomena of spatio-temporal coupling (and the associated power-law reduced distributions of fluctuating observable quantities) are explained in terms of self-organized criticality (SOC) \cite{Bak}. SOC is a paradigmatic concept describing the general tendency of complex driven dissipative dynamical systems to generate power-laws in response to an external driving. Here we witness a different situation according to which the dynamical system is Hamiltonian, yet it generates power-laws through the self-organization of clusters of coupled nonlinear oscillators. The relaxations in this system are multi-scale and mediated by a stripy ordering permitting long-distance jumps of the wave field with both a jump-length and waiting-time statistics. Arguably we may consider these relaxation events as the analog SOC avalanches. According to Eq.~(\ref{JumpL}) these ``avalanches" will be characterized by a size distribution 
\begin{equation}
\chi (\Delta n) \propto  \Delta n^{-(1+\mu)},
\label{Size-D} 
\end{equation}
which corresponds to the elusive {\it gray swans} in the vocabulary of Ref. \cite{PRE2018}. The gray swans \cite{PRE2018} are a specific type of large-amplitude events \cite{Sornette_Ext} that occur at the border of localization-delocalization in complex systems. Tuning $\mu$ to its lower limit at $\mu = 1$ (i.e., $s=1/2$), one also obtains $\chi (\Delta n) \propto  \Delta n^{-2}$. This behavior has been found numerically in Ref. \cite{Sornette_PRL} through a study of coupled nonlinear oscillators with a phase-space instability. At contrast to SOC, the onset of relaxation dynamics in the nonlinear system~(\ref{4s+}) is {\it not} thresholded in terms of the $\beta$ value; at least, in the quantum domain. In this regard, we might also speculate that the occurrence of stripes in an QNLSE system with $s < 1$ is an example of a self-organization {\it without} criticality. This type of dynamical phenomena has attracted some attention in the literature recently \cite{Asch2018}.   

\subsection{Hole superconductivity}

The idea that the localization-delocalization transition in QNLSE~(\ref{1}) occurs through coupling among quantum $s$-waves and the associated stripy ordering finds an interesting parallel in the theory of the superconductivity in some unconventional superconductors such as, for instance, self-assembling organic polymers and copper-oxide compounds. It has been discussed \cite{PRB02,Emery,Cho} that these superconductors are {\it not} described by the traditional picture of Bardeen-Cooper-Schrieffer (BCS) superconductivity in regular crystals \cite{Cooper} in that the superconducting quantum condensate in this type of materials is due to the pairing among the hole states and not the electron states. In fact, the cuprate superconductors consist of parallel planes of copper and oxygen atoms arranged in a square grid. The copper-oxygen planes are separated by the layered atoms of other elements, which may absorb electrons from the copper sites, leaving positively charged holes behind (Ref. \cite{Cho}; references therein). Then the integral system operates like a multilayer field-effect transistor with the conducting copper-oxygen planes confined between the charge-absorbing insulating substrates. 

The key issue about the holes is that they have negative mass $m^* = -|m^*| < 0$, leading to the imaginary fractal tunneling length $l^* \sim i\hbar /\sqrt{2|m^*| T}$. This means a periodic charge oscillation $\sim \exp (-ix\sqrt{2|m^*| T}/\hbar)$ along the position coordinate $x$ in the copper-oxygen lattice, i.e., a ``stripe." It has been shown \cite{PRB02} based on a fracton model for hole-hole interactions that the stripy ordering in cuprate superconductors might result from the self-organization of the conducting system to a thermodynamically profitable one-dimensional charge distribution, and that the generation of the stripes was a mechanism by which the superconducting condensate may flow across the complex molecular structure of the cuprates without resistance. 

Here, we add value to this discussion by proposing that the stripy ordering in quantum nonlinear systems can be understood theoretically based on a QNLSE with subquadratic power nonlinearity. In this paradigm, the loss of Anderson localization in certain quantum nonlinear Schr\"odinger lattices with randomness might be said to be due to the ``superfluidity" of coupled $s$-waves escaping the localization domain along the stripes. One might also speculate an interesting parallel between the $s$-waves and the hole states, as well as the idea that the localization-delocalization transition in QNLSE~(\ref{1}) occurs as a result of the self-organization of a lattice gas of interacting unstable modes for $s \geq 1/2$. The implication is that this type of nonlinearity introduces a dynamical system feedback mechanism \cite{Sornette} prompting the coupled states of counter-propagating quantum $s$-waves to arrange themselves into a one-dimensional ordered structure, hence a phase-transitionlike behavior despite the underlying disorder.   

\subsection{The competition between discontinuity and nonlocality} 

Next we address the subtle difference between the quantum nonlocality$-$which allows the components of quantum $s$-waves to tunnel under the topological barriers in mapping space$-$and the kinetic nonlocality, which permits the coupled states of quantum $s$-waves to perform long-distance jumps along the stripes. The co-existence between these two dynamical processes in a QNLSE system is accounted for by the matching condition $\mu = 2s$, which, together with the requirement that the L\'evy flight trajectories must be path-connected, implies that the transport is possible for all $1/2 \leq s \leq 1$. The latter result is in marked contrast with the classical situation, according to which the transport to long distances is only possible for $s=1$ (see Table~I). Note, in this regard, that the path-connectedness \cite{Nash} of the trajectories in {\it space} is a consequence of the continuity of the random walk process in {\it time} \cite{Bouchaud,CTRW}. 

Finally, the matching condition $\mu = 2s$ demonstrates that the rate of quantum excitation in Eq.~(\ref{Rate}) scales with the system size as the number density of quantum jumps to the distance $\Delta n$. Indeed, it is the density of the jumps that defines dynamically the mode excitation rate, as Fermi's golden rule would imply.

All in all, we might conclude that the quantum transport in QNLSE~(\ref{1}) is an elegant and delicate compromise between quantum tunneling and flights, and that the quantum system can overcome the intrinsic discontinuities in the mapping space by organizing itself into a nonequilibrium dynamical state with a stripy order.  

\subsection{The competition between trappings and flights} 

Our next point here concerns the {\it bifractional} form of the basic kinetic Eq.~(\ref{FDEL}), which involves both the fractional Riemann-Liouville derivative over time, $_{0}\mathrm{D}_t^{1-\alpha}$, and the fractional Riesz derivative $\nabla^{\mu}$ over the space coordinate $\Delta n$. The implication is that the transport process proceeds as a {\it competition} \cite{Klafter2004,UFN} between long-time trapping phenomena with an exit-time statistics$-$introduced by multiple clusters in phase space and the attracting dynamics of the Lennard-Jones type$-$and the L\'evy flights of the coupled states along the $\Delta n$ axis. The overall effect this competition has on phase space transport is subdiffusive scaling in Eq.~(\ref{Sub}); where, the $s$ value is limited to the interval $1/2 \leq s \leq 1$. Setting the exponent $s$ to 1, one arrives at the elusive ``half-diffusion," i.e., $(\Delta n)^2 \propto t^{1/2}$ \cite{PRE17}. On the other extreme, one encounters the limiting value $s=1/2$ corresponding to Cauchy flights and the Hilbert transform operator in Eq.~(\ref{Hilbert}). The Cauchy limit leads to a two-thirds law, i.e., $(\Delta n)^2 \propto t^{2/3}$. At this point, one sees that a stronger transport is found for the correspondingly smaller values of $s$, for which the nonlocal features are generally more pronounced. 

Curiously, the L\'evy flights do {\it not} introduce superdiffusion in phase space (as perhaps they would in the absence of trapping), but on the positive side they ensure that there is unlimited transport for all $1/2 \leq s < 1$, contrary to the classical transport case \cite{PRE14,DNC}. 

Last but not least, the subdiffusion in QNLSE~(\ref{1}) is claimed in the presence of nonlocality$-$counter-intutive, if not substantiated by the kinetic Eq.~(\ref{FDEL}), with properly balanced time- and phase-space derivatives.

\section{The diffusion paradox}

If one is ambitious and wants to reach beyond the condition $s \geq 1/2$, then one may try to tune the $s$ value to zero by allowing $s\rightarrow+0$, which is the absolute bound on the exponent $s$ in QNLSE~(\ref{1}). Then based on the scaling law in Eq.~(\ref{Sub}) one would argue that the transport is {\it diffusive}, i.e., $(\Delta n)^2 \propto t$. This sounds like a paradox, since the limit $s\rightarrow +0$ corresponds to the linear model, for which an Anderson localization would be the case. 

The paradox is solved by demonstrating that the diffusion coefficient vanishes for $s\rightarrow +0$, i.e., there is in fact a diffusive scaling coming up, but the diffusion flux is {\it zero}, implying that there is {\it no} transport in real terms.  

The demonstration refers to the topology of the Cayley trees as of Sec. III above. Letting $s\rightarrow +0$, one sees that the bonds with the self-interference mark (i.e., the Cantor sets in Fig.~1) are none other than the empty sets (since their fractal dimension is exactly {\it zero} in this limit). That means that the Cayley trees defined by the dynamical Eq.~(\ref{4s+}) loose two out of three their bonds at each node, leaving one (and only one) isolated bond behind. In our Fig.~1 this would be the bond connecting the node $k$ to $k_3$, which corresponds to the operator $\hat a_{m_3}$. 

The fact that the non-vanishing bonds are disconnected from other similar bonds would imply that the Cayley trees have lost their structural identity in terms of the coordination number. To this end, they are not even {\it trees} anymore, if not a collection of isolated bonds lying here and there in the graph space. As the topology of a tree has been relaxed in the graph space, also relaxed will be any eventual long-range dependences introduced by such trees in correspondence with the branching process in Eq.~(\ref{4s+}). In particular, the fact that the node $k$ appears to be connected just to $k_3$ (and not to any other node around) becomes a matter of choice. With the deteriorating Cayley trees for $s\rightarrow +0$, the Cayley forest at each node $k$ becomes a collection of randomly oriented bonds, and their respective lengths $k - k_3$ constitute a set of random numbers centered around $k$.  

The net result is that any eventual transport in the limit $s\rightarrow +0$ would be fully {\it uncorrelated} in the long run (both in time and in wave number space). This corresponds to a Gaussian diffusion process, i.e., the familiar Brownian random walk \cite{Klafter,Kampen}. Then it is the particularity of the Anderson problem that this transport process occurs with {\it zero} diffusion coefficient, as we now proceed to show.   

The key step is to consider the series expansion in Eq.~(\ref{Expan}) as a sum of random variables obeying the conditions of the central limit theorem \cite{Gnedenko,Kampen}. Then this theorem tells us that the probability distributions of these sums$-$obtained statistically at the different positions $n$$-$will be Gaussian (or ``normal"). Self-consistently, each of these probability distributions will be none other than the probability density of the wave field itself and will, therefore, be given by $|\hat\psi_n|^{2} =\hat\psi_n^{\dag}\hat\psi_n$. This is said, we can represent the probability density 
\begin{equation}
|\hat\psi_n|^{2} = \sum_{m_1,m_2} \hat a^{\dag}_{m_1} \hat a_{m_2} \phi^*_{n,m_1}\phi_{n,m_2}
\label{PD} 
\end{equation} 
as an effective bell function, i.e., 
\begin{equation}
|\hat\psi_n|^{2} \simeq \exp [-\hat b^\dag\hat b\,\tilde{\phi}_n^*\tilde{\phi}_n],
\label{Gaussian} 
\end{equation} 
where we have also introduced the effective ``normal" creation-annihilation operators $\hat b^\dag$ and $\hat b$; as well as the associated ``effective-medium" complex functions $\tilde{\phi}_n^*$ and $\tilde{\phi}_n$, which may depend on the coordinate $n$ in general. 

Focusing on the bell function in Eq.~(\ref{Gaussian}), one may use the Gaussian representation 
\begin{equation}
|\hat\psi_n|^{2s} \equiv (|\hat\psi_n|^{2})^s \simeq \exp [-s\hat b^\dag\hat b\tilde{\phi}_n^*\tilde{\phi}_n]
\label{GauRe} 
\end{equation}
to define the power $s$ of the probability density $|\hat\psi_n|^{2}$. Note that the representation in Eq.~(\ref{GauRe}) is {\it exact} within the range of validity of the Gaussian central limit theorem. This representation is applied in the limit $s\rightarrow +0$ for which Eq.~(\ref{Gaussian}) holds, making it possible to circumvent the use of the backbone map and other topological approximations alike, as soon as the Gaussian diffusion is considered. Taylor expanding the exponential function for $s\rightarrow +0$, we have $|\hat\psi_n|^{2s} \simeq 1 - s\hat b^\dag\hat b\tilde{\phi}_n^*\tilde{\phi}_n$. Upon substituted in QNLSE~(\ref{1}) this yields  
\begin{equation}
i\hbar\frac{\partial\hat\psi_n}{\partial t} = \hat{H}_L^\prime \hat\psi_n - s \beta \hat b^\dag\hat b [\tilde{\phi}_n^*\tilde{\phi}_n] \hat\psi_n,
\label{1+} 
\end{equation}
where $\hat{H}_L^\prime = \hat{H}_L + \beta$. If we introduce $\omega_k^\prime = \omega_k + \beta$, we may write for the eigenstates $\hat H_L^\prime \phi_{n,k} = \omega_k^\prime\phi_{n,k}$. Confident on Eq.~(\ref{Expan}), we multiply both sides of Eq.~(\ref{1+}) by $\phi_{n,k}^*$, then sum over $n$ using the orthonormality condition in Eq.~(\ref{Orton}). The end result is the following system of equations 
\begin{equation}
i\dot{\hat a}_k - (\omega_k^\prime - s\beta \hat b^\dag\hat b V_{k, k}) \hat a_k = -s\beta \hat b^\dag\hat b \sum_{m\ne k} V_{k, m} \hat a_{m},
\label{4s+G} 
\end{equation}
in which the coefficients
\begin{equation}
V_{k, m} = \sum_{n} [\tilde{\phi}_n^*\tilde{\phi}_n] \phi^*_{n,k}\phi_{n,m}
\label{5s+G} 
\end{equation}
characterize the overlap structure of the wave field for $s\rightarrow +0$. Equation~(\ref{4s+G}) is a {\it linear} dynamical equation and represents a system of forced {linear} oscillators, where the forcing term at the node $k$ is defined by 
\begin{equation}
F_k \simeq -s\beta \hat b^\dag\hat b \sum_{m\ne k} V_{k, m} \hat a_{m}
\label{Forcing_k} 
\end{equation}
and is due to the presence of multiple decoupled oscillators in the surrounding space. For the large number of the oscillators, the forcing term in Eq.~(\ref{Forcing_k}) acts as a Gaussian white noise term in the limit $t\rightarrow+\infty$. At this point, Eq.~(\ref{4s+G}) can conveniently be considered as a system of Langevin equations with the effective-medium white noise term, whose amplitude, $F_k$, depends parametrically on the index $s$. As is well known \cite{Kampen}, the Langevin equations with a Gaussian white noise give rise to a stochastic walk process of the diffusion type, and this supports the observation above that the scaling in Eq.~(\ref{Sub}) is diffusive for $s\rightarrow +0$. On the other hand, the corresponding diffusion coefficient is none other than the intensity of the noise, i.e., 
\begin{equation}
D(s) \simeq s^2 \beta^2 |\hat b^\dag\hat b|^2 \sum_{m_1\ne k,\,m_2 \ne k} V_{k, m_1}^* V_{k, m_2} \hat a_{m_1}^\dag \hat a_{m_2},
\label{Diff_C} 
\end{equation}
and this behaves for the small $s\rightarrow +0$ as $D(s) \propto s^2$. 

The end result is that the diffusion coefficient vanishes (as a square of $s$) in linear random lattices, giving rise to the phenomena of Anderson localization through the loss of long-range correlation in a system of forced, weakly interacting oscillators. All in all, this brings us to the celebrated {\it Absence of Diffusion}, as it has been intimated by Anderson in his seminal work in Ref. \cite{And}. 

\section{Summary and Conclusions}

In the present work, we have analyzed the defining conditions permitting the spreading of an initially localized wave packet in one-dimensional discrete quantum nonlinear Schr\"odinger lattices with disorder and algebraic power nonlinearity. Our results for quantum and classical lattices differ considerably. For classical lattices, we have found through previous investigations \cite{EPL,PRE14,DNC} that an unlimited spreading is only possible in the quadratic power case, and that the phenomenon is {\it thresholded} in that there exists a critical strength of nonlinear interaction, below which the nonlinear field is Anderson localized similarly to the linear field. No unlimited transport of the wave function has been found for subquadratic power nonlinearities as a result of the topological restrictions in the graph space. 

For quantum lattices, we predict an unlimited spreading for both the quadratic and to a certain extent subquadratic power nonlinearities$-$in contrast to the corresponding classical lattices. Moreover, the phenomenon is {\it not} thresholded, i.e., no critical strength of nonlinear interaction comes into play. The explanation lies in the realm of quantum tunneling processes permitting sub-barrier propagation of the wave function, hence unlimited transport in regimes otherwise inaccessible for the classical field. 

Our analysis indicates that the relevant parameter defining the transition from localization to an unlimited spreading is the exponent $s$ of the power-law, which appears in the nonlinear term, $\sim \beta |\hat\psi_n|^{2s}\hat\psi_n$. Then for the quantum lattices we find that an unlimited transport to long distances is possible for all $1/2 \leq s\leq 1$; whereas for the classical lattices it would only be possible for $s = 1$, but not for $s < 1$. Finally, for $0 < s < 1/2$, the quantum field is Anderson localized despite the nonlinearities present. These findings are compiled in Table~I, where one also finds a summary of dynamical properties behind.    

More generally and more importantly, we have devised analytical methods enabling one to tackle algeabraic power nonlinearities in much the same way as the familiar quadratic nonlinearity. Those methods have involved the multinomial theorem \cite{Stegun} jointly with the fopmalism of Diophantine equations and some mapping procedure on a Calyley tree. In that regard, we could argue that the nonlinear Anderson problem was a topological problem of connectedness \cite{Nash} in wave number space mediated by quantum nonlocality properties. 

In the discussion above we have argued that the stochasticity parameter $K = \Delta\omega_{\rm NL} / \delta \omega$ was dynamic in that it would naturally increase while spreading if the nonlinearity is subquadratic (i.e., $s < 1$). That is, the chaos is {\it self-reinforcing} for $s < 1$. In that case $K$ involves a dependence on the number of states through $K\simeq \beta (\Delta n)^{1-s}$. So, if $K$ is large for some $t=0$, then it will be getting even larger for $t>0$, thus dynamically improving the condition $K\gg 1$ for which a kinetic model of the Fokker-Planck type (whatever ordinary or fractional) may be introduced. If $s=1$, then $K\simeq \beta$ is an invariant of the spreading process, a property already discussed in Refs. \cite{PS,PRE14}. 

It is understood that the quantum transport is much faster than the classical estimates would predict. Indeed, given a QNLSE with the quadratic power nonlinearity, we have found using Fermi's golden rule $(\Delta n)^2 \propto t^{1/2}$; whereas the classical estimates lead to $(\Delta n)^2 \propto t^{2/5}$ in the chaotic domain, and to $(\Delta n)^2 \propto t^{1/3}$ in the pseudochaotic domain \cite{EPL,PRE14}. In the latter case one recovers the one-third law in Eq.~(\ref{Binom3}). We associate the observed deviations with the fact that the topological constraints for quantum transport are less demanding than in the classical transport case; in particular, the quantum field is allowed to tunnel along the disconnected bonds in the graph space, as well as interfere with itself under the topological barriers to form quasiparticle states$-$the processes that are strictly forbidden in classical lattices.   

Another particularity of the quantum case is the mechanism of the transport: 

For $s= 1$, the transport (both quantum and classical) is absolutely dominated by long-time trapping phenomena and is non-Markovian in nature. The trappings occur because the nonlinear interactions act as to introduce an attracting potential in wave number space favoring some kind of stickiness phenomena \cite{Report} among the unstable modes. By examining the structure of the nonlinear term one finds that the attracting potential is none other than the celebrated Lennard-Jones potential giving rise to the formation of multiple clusters$-$or ``molecules"$-$in wave number space, which could effectively reduce the transport below the expected diffusion values.  

It is worth stressing here that the non-Markovian properties arise naturally through dynamics via the action of the Lennard-Jones potential causing attraction between the unstable modes. It is due to this attraction that the actual transport is much slower than a diffusive one. Behind the subdiffusive character of the spreading is the nonlinear interaction between the modes; in particular, the ``$(2s+2)$"-wave interaction in Eq.~(\ref{4s+}) generates a waiting-time distribution with the divergent mean, enabling non-Markovian dependencies in Eq.~(\ref{FDEL}). 

For $1/2 \leq s < 1$, there is an additional mechanism coming into play, and this is effective in quantum lattices only$-$not the classical ones. It uses the possibility that some components of the wave field, the so-called $s$-waves, may constructively interfere under the topological barriers caused by multiple discontinuities in the operator space, and by doing so can produce coupled states with zero overall momentum$-$similar to the hole-hole pairing processes in complex superconducting materials \cite{PRB02,Cho}. Then these pairwise coupled states can propagate to long distances on L\'evy flights by jumping over the discontinuities in the mapping space. The process is favored in the presence of a simultaneous long-range one-dimensional ordering dubbed stripy ordering. For classical lattices, this mechanism is forbidden, because the wave function cannot penetrate under the barrier to meet its pairing counterpart. It is due to this pairing mechanism that the quantum transport for $1/2 \leq s < 1$ is at all possible.    

It is worth noting here that the lower bound on the exponent $s$, i.e., $s = 1/2$, is dictated by the {connectedness} of the L\`evy flight trajectories through the condition $d_f = 2s \geq 1$, where $d_f$ is the fractal dimension of the flight. Also it is dictated by the connectedness of the stripes in wave number space navigating the escape of the wave field to long distances. We have seen in the above that these connectedness properties being essentially the {\it topological} properties of a wave field as complex system could be translated in terms of the {continuity} of the transport process in {\it time}. This may be seen in the fact that the exponents of both the waiting-time distribution in Eq.~(\ref{WaitingT}) and the jump-length distribution in Eq.~(\ref{JumpL}) are categorized by the same parameter $s \leq 1$ and in this sense are {\it not} independent. 

In a basic physics perspective, the L\'evy flights introduce important {\it nonlocal} features into the dynamics and by doing so contest Fick's first law \cite{Sokolov,Fick} that fluxes are generated by local gradients. The comprehension of nonlocality of quantum transport has led us to a kinetic description based on the theoretical scheme of continuous time random walks \cite{Klafter,CTRW}. When applied to the QNLSE problem in Eq.~(\ref{1}), this scheme assumes that the transport occurs as a competition between multiple trapping phenomena due to the binding effect of the phase space clusters, on the one hand, and the L\'evy flights of the coupled states to long distances along the stripes, on the other hand. For $t\rightarrow +\infty$, this scheme leads to a description in terms of a time- and L\'evy-fractional kinetic Eq.~(\ref{FDEL}), using the Riemann-Liouville derivative $_{0}\mathrm{D}_t^{1-\alpha}$ for the trapping phenomena, and the Riesz fractional derivative $\nabla^{\mu}$ for the flights. We note in this regard that the fractal dimension of the L\'evy flights is determined by the exponent of the algebraic power nonlinearity and is equal to $2s$ exactly, hence the connectedness restriction $2s \geq 1$. For $s=1$, the L\'evy flights do not occur. Controlling this case will be the standard Gaussian central limit theorem \cite{Gnedenko,Kampen}, as the matching condition in Eq.~(\ref{MuS}) shows. That means that there will be a characteristic jump length along the lattice (which defines the half-width of the Gaussian bell), and not a heavy-tailed distribution of these, at contrast to Eq.~(\ref{JumpL}). As a consequence, the Riesz operator $\nabla^\mu$ in the basic kinetic Eq.~(\ref{FDEL}) must be replaced by a Laplacian operator $\Delta = \nabla^2$. This recovers the next-neighbor random walk model already discussed in Ref. \cite{PRE17}. 

The fact that the exponents $\mu$ and $\alpha = \mu / (\mu + 2)$ prove to be related to each other means that there is an important spatio-temporal coupling at play governing the escape of the nonlinear field to infinity. When seen from a perspective of the competing multiple trappings and flights this escape process to infinity (and the associated destruction of Anderson localization in quantum nonlinear Schr\"odinger lattices) turns out to obey the bifractional kinetic equation in Eq.~(\ref{FDEL}).   
 
Our next point here concerns the fact that the resulting transport process incorporating both the waiting-time statistics and L\'evy flights of the coupled states is {\it subdiffusive} despite the nonlocalities present. The implication is that the binding effect of phase space clusters is quite strong in that it overcomes the nonlocality effect due to the stripes. As a consequence, the actual transport level is below the diffusionlike levels.   

By tuning the exponent $s$ to zero we have demonstrated that the limit $s\rightarrow +0$ corresponds to a {\it diffusive} scaling in a system of weakly interacting, linear forced oscillators. Because of the lack of connectedness in the limit $s\rightarrow +0$, the diffusion coefficient (i.e., the flux of the field) goes to zero as $D(s) \propto s^2$. One sees that the flux of the field vanishes, as soon as the diffusion limit is achieved. This kind of paradoxical halt of diffusion is, in fact, the phenomenon of Anderson localization \cite{And}. 

Experimentally, detailed identification of the driving mechanisms leading to the destruction of Anderson localization in quantum nonlinear lattices with disorder is in its infancy. Beyond validation of theoretical models, the future of the field lies in the development of accurate numerical simulation tools. A few important milestones have been achieved recently in the activities of Flach and co-workers (e.g., Ref. \cite{Flach_89}), who used a simplification of QNLSE~(\ref{1}) based on the Hubbard model, with the structure of the nonlinear term corresponding to our $s=1$. The results support the idea that the asymptotic transport is subdiffusive, with the transport exponent complying with a ``half-diffusion" process \cite{PRE17}. The latter process is a partial case of the more general scaling law in Eq.~(\ref{Sub}). 

Extending the numerical simulations to an arbitrary $0 < s < 1$ is not at all trivial. The inclusion of the fractional values $0 < s < 1$ implies that the wave field may partially annihilate along some bonds in the graph space. Then the integral picture of the interactions is more like an overlap among the fractional number $2s + 2$ of the different waves, rather than the familiar 4-wave interaction dynamics \cite{Iomin}. On top of this, the partially annihilated modes may stick together to produce coupled states, and these may propagate along the lattice on L\'evy flights introducing the important features of nonlocality into the transport. We note in this regard that the nonlocal features are absent in the quadratic power case for which the transport is local in the sense of Fick's first law and the Gaussian central limit theorem. Another complication comes with the fact that the numerical simulation tool should be able to capture the signatures of the asymptotic transport in the limit $t\rightarrow+\infty$. This may result in certain numerical controversies already discussed in Refs. \cite{EPL,PRE14}. 

All in all, we expect the eventual numerical simulation task being both an intricate and challenging problem, which naturally constitutes an important subject for future investigations. We consider this task as posing the frontier for the present study.

\acknowledgments
This study has been carried out within the framework of the EUROfusion Consortium and has received funding from the Euratom research and training programme 2014-2018 under Grant agreement No. 633053 for the project AWP17-ENR-ENEA-10. A.V.M. gratefully acknowledges the hospitality and partial support at the International Space Science Institute (ISSI) at Bern, Switzerland, where this paper was written. Also A.V.M. thanks D. Sornette for sharing insights into the topic of extreme events, and for highlighting the study in Ref. \cite{Sornette_PRL}. The work of A.I. was supported by the Israel Science Foundation (Grant ISF-931/16). The views and opinions expressed herein do not necessarily reflect those of the European Commission. 

%
%
%
%


\begin{thebibliography}{}
\bibitem{DNA}
By the term ``DNA" we mean the familiar carrier of genetic information. Here used in a figurative sense.
\end{thebibliography}

\end{document}